\documentstyle[11pt,aaspptwo]{article} 
\def\fnsiz{\footnotesize}
\def\ts{\thinspace}
\newlength{\zwidth}
\newcommand{\as}{\mbox{$''$}}
\newcommand{\Brg}{\mbox{Br{\fnsiz $\gamma$}}}
\newcommand{\dg}{\mbox{$^{\circ}$}}
\newcommand{\ebv}{\mbox{$E(B-V)$}}
\newcommand{\ebvbd}{\mbox{$E(B-V)_{\rm BD}$}}
\newcommand{\ebvgal}{\mbox{$E(B-V)_{\rm gal}$}}
\newcommand{\ebvuv}{\mbox{$E(B-V)_{\rm UV}$}}
\newcommand{\fion}[2]{\mbox{\rm [{#1}\ts{\fnsiz {#2}}]}} 
\newcommand{\Foha}{\mbox{$F_{{\rm H}{\alpha},0}$}}
\newcommand{\FIR}{\mbox{\em FIR\/}}
\newcommand{\FU}{\mbox{erg cm$^{-2}$ s$^{-1}$}}
\newcommand{\FUA}{\mbox{erg cm$^{-2}$ s$^{-1}$ \AA$^{-1}$}}
\newcommand{\gapeq}{\mbox{$~\stackrel{\scriptstyle >}{\scriptstyle \sim}~$}}
\newcommand{\HI}{\mbox {H\thinspace{\fnsiz I}}}
\newcommand{\HII}{\mbox {H\thinspace{\fnsiz II}}}
\newcommand{\Hline}[1]{\mbox{H{\fnsiz {#1}}}}
\newcommand{\Halpha}{\Hline{\mbox{$\alpha$}}}
\newcommand{\kms}{\mbox{km\ts s$^{-1}$}}
\newcommand{\lapeq}{\mbox{$~\stackrel{\scriptstyle <}{\scriptstyle \sim}~$}}
\newcommand{\msfae}{\mbox{\rm MSFA$_e$}}
\newcommand{\muv}{\mbox{$m_{220}$}}
\newcommand{\Muv}{\mbox{$M_{220}$}}
\newcommand{\Msun}{\mbox{$M_\odot$}}
\newcommand{\pino}{\parindent=0mm}

\newcommand{\Wfo}{\mbox{$W_{50}$}}
\newcommand{\zsp}{\mbox{\hspace{\zwidth}}}

\received{25 May, 1995}
\revised{24 Aug, 1995}
\accepted{28 Aug, 1995}

\begin{document}
\parskip=1ex

\title{Starbursts and Star Clusters in the Ultraviolet
\footnote{Based on observations with the NASA/ESA {\em Hubble Space
Telescope\/} obtained at the Space Telescope Science Institute, which is
operated by the Association of Universities for Research in Astronomy,
Inc., under NASA contract NAS5-26555.}}

\author{G.R.\ Meurer\footnote{e-mail contact:
meurer@poutine.pha.jhu.edu.}, T.M.\ Heckman}
\affil{Department of Physics and Astronomy, \\
The Johns Hopkins University }

\author{C.\ Leitherer, A.\ Kinney}
\affil{Space Telescope Science Institute}

\author{C.\ Robert}
\affil{D\'epartement de Physique, Universit\'e Laval, \\
and Observatoire du Mont M\'egantic}

\and

\author{D.R.\ Garnett\footnote{Hubble Fellow.}}
\affil{Astronomy Department\\
University of Minnesota}
\vskip 10mm

\centerline{Accepted for publication in {\em The Astronomical Journal}}

\clearpage

\begin{abstract}
We present ultraviolet (UV) images of nine starburst galaxies obtained
with the {\em Hubble Space Telescope\/} using the {\em Faint Object
Camera}.  The galaxies range in morphology from blue compact dwarfs to
ultra-luminous merging far-infrared galaxies.  Our data combined with
new and archival UV spectroscopy and far-infrared fluxes allow us to
dissect the anatomy of starbursts in terms of the distributions of
stars, star clusters and dust.

The overall morphology of starbursts is highly irregular, even after
excluding compact sources (clusters and resolved stars).  The
irregularity is seen both in the isophotes and the surface brightness
profiles.  In most cases the latter can not be characterized by either
exponential or $R^{0.25}$ profiles.  Most (7/9) starbursts are found
to have similar intrinsic effective surface brightnesses, suggesting
that a negative feedback mechanism is setting an upper limit to the
star formation rate per unit area.  Assuming a continuous star
formation rate and a Salpeter (1955) IMF slope, this surface
brightness corresponds to an areal star formation rate of $0.7\,
\Msun\, {\rm Kpc^{-2}\, yr^{-1}}$ in stars in the mass range of 5 --
100 \Msun.

All starbursts in our sample contain UV bright star clusters
indicating that cluster formation is an important mode of star
formation in starbursts.  On average about 20\%\ of the UV luminosity
comes from these clusters.  The clusters with $M_{220} < -14$ mag, or
super star clusters (SSC) are preferentially found at the very heart
of starbursts; over 90\%\ of the SSCs are found where the underlying
surface brightness is within 1.5 mag arcsec$^{-2}$ of its peak value.
The size of the SSCs in the nearest host galaxies are consistent with
those of Galactic globular clusters.  Our size estimates of more
distant SSCs are likely to be contaminated by neighboring clusters and
the underlying peaked high surface brightness background.  The
luminosity function of SSCs is well represented by a power law
($\phi(L) \propto L^{\alpha}$) with a slope $\alpha \approx -2$.

We find a strong correlation between the far infrared excess and the
UV spectral slope for our sample and other starbursts with archival
data.  The correlation is in the sense that as the UV color becomes
redder, more far-infrared flux is observed relative to the UV flux.
The correlation is well modeled by a geometry where much of their dust
is in a foreground screen near to the starburst, but not by a geometry
of well mixed stars and dust.  Some starbursts have noticeable dust
lanes, or completely obscured ionizing sources, indicating that the
foreground screen is not uniform but must have some patchiness.
Nevertheless, the reddened UV colors observed even in these cases
indicates that the foreground screen has a high covering factor and
can account for a significant fraction of the far-infrared flux.

\end{abstract}

\section{Introduction}\label{s:intro}

Starburst galaxies are currently forming stars at a much higher rate
than the past average.  In the most extreme cases the star formation
rate (SFR) is the highest permissible (i.e.\ SFR $\approx$ gas
mass/dynamical time scale; Heckman, 1994).  The {\em intrinsic\/}
optical signatures of a starburst are a high surface brightness and a
spectrum dominated by strong, narrow, high-excitation emission lines,
a relatively weak continuum that is flat or rising bluewards and weak
absorption features (primarily Balmer series).  These signatures are
often diminished by dust extinction. An important subclass of
starburst galaxies are the far infrared galaxies (FIRGs; Soifer et al.\
1987; Armus et al.\ 1990, hereafter AHM).  These are very dusty systems
in which the dust grains act to redistribute the intrinsic spectrum
from the ultraviolet (henceforth UV) and optical to the far infrared,
where most of the bolometric luminosity is emitted.  Galaxies with
starburst characteristics range in morphology from blue compact dwarfs
(BCDs; Thuan \&\ Martin 1981) to merging systems.  The starburst
itself is often confined to a small portion of a galaxy, frequently
just the very central regions of a large spiral in which case they are
classified as starburst nuclei galaxies (Balzano, 1983). Spectacular
galactic winds are often observed in starburst galaxies ranging in
luminosity from BCDs (e.g.\ Meurer et al.\ 1992, hereafter MFDC; Marlowe
et al.\ 1995) to ultraluminous FIRGs (AHM; Heckman et al.\ 1990).

High mass stars ($m_* > 10$\Msun) especially the hottest, are the
driving engines of starbursts.  Ultimately, they provide most of the
luminosity, are responsible for ionizing the ISM and warming the dust,
and through their stellar winds and supernovae power the galactic
winds.  Although much of what we know about starbursts is from optical
and infrared observations, it is the UV where these stars directly
dominate the spectral energy distribution.

Numerous spectroscopic studies of starbursts using the {\em
International Ultraviolet Explorer}\linebreak (IUE) have been
published.  Recently, Kinney et al.\ (1993, hereafter K93) presented
an atlas of IUE spectra of 143 star forming galaxies, many of them
starbursts.  UV imaging studies of galaxies are somewhat rarer.  Most
previous studies have been obtained either from balloon-born
experiments such as FOCA (e.g.\ Buat et al.\ 1994; Reichen et al.\
1994; Courvoisier et al.\ 1990; Donas et al.\ 1987) or limited
duration space flights such as FAUST (Deharveng et al.\ 1994) and UIT
(e.g.\ Hill et al.\ 1992; Chen et al.\ 1992).  Mostly ``normal''
galaxies were observed by these experiments, at angular resolutions of
$\sim 4''$ to $\sim 3.5'$.  With the launch of the {\em Hubble Space
Telescope\/} (HST) we can now obtain UV images at $\sim$50 mas
(milli-arcsecond) resolution, or physical sizes of $\sim 2$ pc at a
distance of 10 Mpc.  Thus we now have the capability of determining
the detailed structure or ``anatomy'' of a starburst at a wavelength
near where the hot high mass stars dominate.  This is the next best
thing to seeing the intrinsic distribution of ionizing radiation
($\lambda < 912$\AA), which of course is impossible since very little
of it escapes the starburst's ISM.

In this paper we examine the UV anatomy of nine starburst galaxies
imaged by HST.  The UV observation yield important results concerning
the distributions of star formation and dust in starbursts.  The
selection of the sample and data reduction and analysis are discussed
in \S\ref{s:data}.  The properties our observations are sensitive to
and the morphology of each galaxy is discussed in \S\ref{s:morph}.
The global properties of the galaxies (e.g.\ integrated fluxes,
spectral slopes, mean surface brightness) are presented in
\S\ref{s:prop}.  These are heavily affected by dust.  The combination
of UV and far-infrared photometry and UV spectroscopy provide strong
constraints on the dust distrubution.  These are discussed in
\S\ref{ss:red}. The properties after deshrouding the dust distribution
are presented in \S\ref{ss:glob}. To derive physical properties such
as mass, and reddening, we compare UV properties with stellar
population models which are derived from the models of Leitherer \&\
Heckman (1995; hereafter LH) and Bruzual \&\ Charlot (1993). These
models are presented in the appendix.

The sample galaxies all contain UV bright compact objects.  We
identify the brightest of these as star clusters.  Cluster formation
is an ongoing process in star forming galaxies as near as the LMC.
High luminosity ``super'' star clusters (SSCs) have been identified
from the ground in the BCD/amorphous galaxies NGC~1569 (Arp \&\
Sandage 1985) and NGC~1705 (Melnick et al.\ 1985; MFDC).  With the
launch of HST, SSCs are being discovered more frequently in such
galaxies as the amorphous galaxy NGC~1140 (Hunter et al.\ 1994a), the
Wolf-Rayet galaxy He2-10 (Conti \&\ Vacca 1994), the peculiar galaxy
NGC~1275 (Holtzman et al.\ 1992), the merging systems NGC~7252
(Whitmore et al.\ 1993) and NGC~4038/4039 (Whitmore \&\ Schweizer,
1995; hereafter WS), and within the circumnuclear starburst rings
around the Seyfert galaxies NGC~1097 and NGC~6951 (Barth et al.\
1995).  These galaxies can also be classified as starbursts, or as
containing starbursts.  So one of our most important conclusions is
that {\em cluster formation is an important mode of star formation in
starburst galaxies}.  In \S\ref{s:clust} we examine the nature of the
clusters.  In particular we consider their location within the
starbursts, we estimate their sizes, and derive a luminosity function.

In \S\ref{s:summ} we summarize our results and what they tell us about
the anatomy of starbursts.

\section{The Observations}\label{s:data}

\subsection{Sample selection}\label{ss:select}

Our sample was selected from the K93 IUE atlas.  The main selection
criteron was UV brightness.  Since the K93 fluxes are limited by the
$\sim 10'' \times 20''$ IUE aperture, this translates to a lower limit
on the UV surface brightness within this aperture.  The final sample
was selected to cover a wide range in metallicity and luminosity, and
to have low foreground extinction.  Table~\ref{t:prop} summarizes much
of what was previously known about the sample from optical
observations.  The properties include distance, $D$, determined from
the radial velocity, $V_r$, after a Virgocentric flow correction and
assuming $H_0 = 75\, {\rm \kms\, Mpc^{-1}}$ (except for NGC~5253 for
which we adopt the Cepheid distance of Sandage et al., 1994), the
absolute magnitude $M_B$, and the morphology.  See Table~\ref{t:prop}
for details on how these quantities were calculated and the sources of
the data.  Since we are primarily concerned with UV data it is
important to parameterize the amount of extinction due to dust.
Table~\ref{t:redex} gives relevant estimates of reddening, \ebv, and
extinction $A_\lambda$, both from the Milky Way foreground and
internal to the sample galaxies.  Reddening and the internal
extinction are discussed in detail in
\S\ref{ss:red}.

The sample covers a wide range of starburst types from low luminosity,
virtually dust free BCDs, such as IZw18, to dusty FIRGs in merging
systems, like NGC~3690.  Most of the starbursts are located near
the center of the host galaxy.  The clearest exception is NGC~3991, an
Im galaxy with an off-center starburst.  In addition, IZw18 is so
irregular in the optical and radio that it is not clear where the
dynamical center is (Dufour \&\ Hester 1990).  A detailed examination
of the environment of the sample is beyond the scope of this paper.
Nevertheless it is fair to say that many of the galaxies are in groups
where the chances for interactions are high (e.g.\ NGC~3991, NGC~5253,
NGC~7552, and especially NGC~3690) while others appear to be fairly
isolated (e.g.\ NGC~1705).

\subsection{Data Reduction and Analysis}\label{ss:reduc}

Ten images of the nine sample galaxies were obtained with the Faint
Object Camera (FOC) and the F220W filter (and neutral density filter,
F1ND, as required).  Due to a fortuitous scheduling error, two
exposures of NGC~1705 were obtained. The reduced images are reproduced
in Fig.~{1} (plates XXXX).  Table~\ref{t:obslog} summarizes
the observations. The positions correspond to some well defined
object, usually a bright cluster, that we define as the coordinate
system origin.  They should be accurate to $\sim 0.7''$ which is the
sum in quadrature of the 0.6\as\ accuracy of the HST Guide Star
Catalog (Lasker et al.\ 1990; Russell et al.\ 1990) and the error in the
FOC aperture position of 0.24\as\ (Hack \&\ Nota 1994).  However, no
external verification of the astrometry has been made for our frames.
The images were obtained in fine lock mode, before the refurbishment
mission, thus they suffer from spherical aberration.  The largest FOC
format, $512z \times 1024$, was employed, yielding a $22'' \times
22''$ field, and rectangular pixels in the raw image. The initial data
reduction, with the standard HST pipeline processing (Nota et al.\
1993), includes dark-count subtraction, ``de-zooming'' the rectangular
pixels by splitting the flux into two square pixels, calculating the
quantities required for photometric calibration, removal of geometric
distortions, and division by a heavily smoothed (by $\sim 15$ pixels)
flatfield.  The final pixel size is 22.5 mas (Baxter 1993).  Most
further processing was done within IRAF, the Image Reduction and
Analysis Facility.

Pixels in the cores of the brightest clusters in NGC~1705, NGC~3690,
and NGC~7552 have raw data values that exceed the 8-bit word length of
the full format images and thus ``wrap around''.  The NGC~3690 and
NGC~7552 images were fixed by adding appropriate multiples of 256 to
the affected pixels of the raw images and then re-running the pipeline
calibration.  This was not feasible for the NGC~1705 frames because of
saturation near the center of its dominant cluster, NGC1705-1.  They
were not corrected.

The photometric calibration coefficients were adjusted for two
effects.  The first is, the slow deterioration with time of the detector
quantum efficiency.  We calculate a $\sim 3\%$ deterioration relative
to the mean epoch of the calibration observations ($\sim 1990.9$;
Sparks, 1991) from data obtained with the F210M filter by Greenfield
et al., (1993). Secondly, the pipeline calibration observations
employed the $512\times 512$ format, while the $512z\times 1024$
format is actually more sensitive by 25\%\ (Greenfield 1994b).  There
is then a net 22\%\ increase in sensitivity for observations relative
to the pipeline calibration.

The FOC is a photon counting device which has a nonlinear response to
incident flux.  Its linearity behavior depends on the source
distribution.  We corrected for the ``flatfield'', or large scale
length, component of non-linearity using the algorithm of Baxter
(1994a,b), and the appropriate linearity parameter from Nota et al.\
(1993).  The maximum correction factor was arbitrarily limited to 1.7.
On average the linearity correction increases the total flux of our
objects by 7\%\ (the range is 3\%\ to 14\%).  On smaller scales the
correction is more important.  For example at the center of NGC~7552
the mean correction factor is 1.5 in a central $25 \times 25$ box.
Linearity corrections at this level are highly uncertain.  However,
since only a small portion of the galaxy is affected, the total flux
is relatively well defined.

Baxter (1994a,b) finds that the additional non-linearity of point
sources is negligible out to an integrated count rate of $C \sim 4.1$
Hz in a $5\times 5$ aperture in frames that have been corrected for
flatfield non-linearity.  Only the brightest clusters (one
or two objects per galaxy) in NGC~3310, NGC~3690, NGC~4670, NGC~7552,
and Tol1924-416 are beyond this limit.  The flux of NGC~1705-1 was not
estimated from aperture photometry so is not affected by
point-source non-linearity.  Baxter (1994a,b) notes that the FOC
zoomed modes appear to be 10\%\ less sensitive to point sources than
to extended sources, in data that has been corrected for flatfield
non-linearity.

The IUE spectra of K93 provide us with an excellent source to check
our photometry as well as complimentary spectral information.  In
Fig.~{2} and Table~\ref{t:spprop} we compare the FOC and
IUE magnitudes of our sample.  The \muv\ magnitudes are on the
``STMAG'' system which is described in the appendix.  $\muv({\rm
FOC})$ was extracted using a circular aperture matching the area of
the IUE extraction aperture of K93.  $\muv({\rm IUE})$ was extracted
using the IRAF package SYNPHOT to define the combined system
throughput.  Table~\ref{t:spprop} also presents the UV spectral slope
$\beta$ ($f_\lambda(\lambda) \propto \lambda^\beta$ fitted to the IUE
spectra using the continuum windows of Calzetti et al.\ 1994; hereafter
C94). The redleak parameter $RL$ (defined in the appendix) was also
extracted from the spectra, but was always found to be negligible, as
expected ($|RL| \leq 0.03$ mag).  Three of the galaxies have Faint
Object Spectrograph (FOS) spectra ($\lambda\lambda 1205-2320$\AA) data
available which were obtained with a 1\as\ aperture.  These spectra
are discussed fully by Robert et al.\ (1995).  $\beta$ measured from
these spectra are also presented in Table~\ref{t:spprop}.

Figure~{2} shows that the HST and IUE magnitudes agree
very well except for NGC~3690, which forms with IC~694, the merging
system Arp~299. K93 note ``The IUE aperture contains both objects''.
This is impossible since each is about 15\as\ in diameter with their
centers separated by $\sim 23''$ (Wynn-Williams et al., 1991).  The low
IUE flux suggests that the IUE aperture was not centered on NGC~3690.
The large disagreement between the IUE and FOS values of $\beta$
supports this view.  The unweighted mean difference $\langle m_{220}({\rm
FOC}) - m_{220}({\rm IUE}) \rangle = -0.05 \pm 0.05$ mag (after
excluding NGC~3690) is slightly less than zero, perhaps due to the
difficulty in centering the IUE aperture.  The uncertainty is the
standard error of the mean.  The rms about the mean is 0.13, the same
as the average error in the difference, indicating that random errors
dominate the error in the mean.

A point spread function (PSF) is required at various stages in the
analysis.  We employ two empirical F220W PSFs in this study; one made
available to us by D.\ Baxter, the other obtained directly from the
HST data archive.  Most of the data manipulation described below used
the former PSF, while the latter was used primarily as a check.  The
PSFs were corrected for flatfield non-linearity.  Neither perfectly
matches the actual PSF of the images because of the effects of focus
differences due to desorption in the optical telescope assembly and
``breathing'' as the HST makes a day-night transition (Baxter et al.,
1993).  In some images (those of NGC~1705, NGC~3690, NGC~4670, and
Tol1924-416) rings around the brightest sources can clearly be
detected.  These are not intrinsic to the galaxy, but are the faint
PSF wings of the bright source at their center.  For these images the
PSF was magnified to match the observed ring radii and then used in
the image restoration described below.  This is an appropriate
procedure, since to first order a focus change contracts or expands
the location of features in the PSF halo (Baxter et al.\ 1993).  The
magnification factors of $1.03 - 1.07$ indicate focus changes of $\sim
10$\micron.

Figure~{1} shows that the sample galaxies contain numerous
embedded compact sources.  We have attempted to separate this clumpy
structure {\em and associated PSF halos\/} from the smooth structure,
in order to estimate their relative contributions to the total light.
This was done by first restoring (deconvolving) the images using 25
standard iterations of the Richardson-Lucy algorithm (Richardson,
1972; Lucy, 1974).  Point-like sources were found using the DAOFIND
algorithm in IRAF's implementation of the DAOPHOT package (Stetson,
1987).  Note that the DAOPHOT package was not used to measure the
magnitude of the sources, only to find them.  The sources were then
``zapped'' (replaced with a fitted surface plus artificial noise) from
the restored image.  The resultant image was reconvolved with the PSF,
and then smoothed with a $15 \times 15$ pixel median filter to yield
the smooth image.  The clumpy image is the difference between the
original image and the smooth image.

The NGC~1705 images required special attention due to the saturated
core of NGC1705-1.  We removed this source by modeling its strength
from the wings of its radial profile (at $0.45'' \leq r \leq 2.15''$).
NGC1705-1 was assumed to have an intrinsically circularly symmetric
Gaussian profile with a full width at half maximum, $\Wfo = 46$
mas\footnote{This is an early estimate of the cluster size determined
from the PC images of O'Connell et al.\ (1994).  A better estimate is
given in \S\ref{ss:sizes}.}.  The model yields $\muv = 12.72$ mag for
NGC1705-1, to within 0.01 mag between the two frames. Since the core
of the images have been excluded, this measurement is not affected by
point source non-linearity. The model cluster was convolved with the
PSF and subtracted from the images.  The residuals in the core region
were zapped.  Figure~{3}a,b (plate XXXX) shows the central
region of NGC~1705a frame before and after the removal of NGC1705-1.
The smooth-clumpy separation was then performed on the images with
NGC1705-1 removed.

Figure~{4} schematically shows the results of the smooth --
clumpy separation.  The panels show the isophotes from the smooth
images superimposed with the objects found by DAOFIND (and NGC1705-1).
The symbol size indicates the magnitude of the objects as determined
from aperture photometry of the linearity corrected images, using a
circular aperture of radius $r=3$ pixels, with ``sky'' taken as the
mode in an annulus of $r = 4$ to 7 pixels.  The aperture
correction to total flux of this aperture is --2.26 mag, as determined
from both the primary and secondary PSFs.  We adopt a final aperture
correction of --2.36 mag to account for the apparent decreased
sensitivity to point sources discussed above.

Surface brightness, $\mu$, profiles were extracted from each frame and
are illustrated in Fig.~{5}.  The extraction procedure was
complicated because of the highly irregular morphology of the targets;
many are club shaped, and typically the isophote centers, and shapes
vary with isophote level.  Our approach is to extract the mean number
of counts in elliptical annuli and convert to $\mu$.  Bad areas (the
distorted edges of the frame, the FOC occulting fingers and a bad
scratch) are masked out of the analysis.  The ellipse parameters are
central coordinates, $\Delta\alpha$, $\Delta\delta$ (relative to
coordinates in Table~\ref{t:obslog}); axial ratio, $a/b$; and position
angle, $\phi$, which are a function of semi-major axis length, $a$.
We hold these parameters constant beyond some outer $a = a_o$, and
inwards of some inner $a= a_i$.  For $a_i < a \leq a_o$ the parameters
are set to vary linearly with $a$ between the values at $a_i$ and
$a_o$. The adopted ellipse parameters at $a_i$ and $a_o$ are given in
Table~\ref{t:ellpar}.  The parameters at $a_o$, meant to be
representative of the outer isophotes, were determined using a moment
analysis technique (Meurer et al.\ 1994) applied to the smooth
component images.  The exceptions were NGC~5253 and NGC~3310.  Due to
their size, few complete isophotes are found; eye estimates of the
parameters were used instead.  At $a_i$, the parameters are set to be
circular, centered near the $\Delta\alpha$, $\Delta\delta$ at $a_i$
determined from the moment analysis.  Circular parameters are adopted
because PSF blurring tends to make the isophotes rounder towards the
target centers, especially if they are highly concentrated.  Often
there is a bright compact source near the center determined from the
moment analysis.  In these cases $\Delta\alpha_i$, $\Delta\delta_i$
were adjusted to coincide with this source.

The $\mu$ profiles were extracted out to $a_{\rm max}$, where the
surface brightness is twice the uncertainty in the background level.
For NGC~3310, $a_{\rm max}$ is where less than half the area of the
annuli are on un-masked portions of the frame.  $\mu$ profiles were
extracted from both the total images and the smooth component images.
The annuli widths were 50, and 350 mas, respectively. NGC~3690 shows
three distinct ``blobs'' (see below) in the FOC images and the
profiles of each of these were measured separately.  Some properties
of the targets extracted from the surface photometry are presented in
Table~\ref{t:surfres}.  These properties include $a_{\rm max}$; $a_e$,
the semi-major axis length containing half the light; $m_T$, the total
\muv\ found by reintegrating the $\mu$ profile out to $a_{\rm max}$;
$\mu_e$, the surface brightness within $a_e$; and $f_{\rm clumpy}$ the
fraction of the flux coming from the clumpy image.  These quantities
are limited by the FOC field of view.  In most cases
the sources are fairly well contained within the detector field of
view, although some have low surface brightness features extending
beyond the frame (NGC~4670, NGC~3690, NGC~3991).  Only in two cases,
NGC~3310 and NGC~5253, do the UV distributions extend beyond the frame
at high surface brightness.  Except in these two cases, the background
count rates of the frames are very similar, suggesting that ``sky''
and not diffuse emission from the host galaxies dominates the
background.

\section{UV morphology}\label{s:morph}

\subsection{What the observations are sensitive to}

The optical spectra of the sample galaxies are all dominated by bright
narrow emission lines, as are most of the galaxies in the K93 atlas.
This indicates that they contain an ionizing population of stars.  We
assume it is this population, that is the stars formed at the same
epoch, that dominate the FOC and IUE observations. This ionizing
population assumption provides a strong constraint on the intrinsic UV
properties of the sample, particularly on the UV spectral slope
$\beta$ (appendix). As shown in the appendix, the F220W filter is well
suited to observing high mass stars.  For a Salpeter (1955)
IMF, which we adopt, extending up to $m_u = 30 - 120$~\Msun\ and
continuous star formation for 10 Myr, the median  stellar
mass our observations are sensitive to is $\sim$20~\Msun, and stars
with $m_* < 5$ \Msun\ provide negligible flux.  Although 20~\Msun\
stars contribute to the ionizing flux, the emission shortwards of
912\AA\ primarily originates in stars with $m_* \gapeq 50$~\Msun.
Likewise, these most massive stars are individually the brightest in
\Muv\ but are insufficient in number to contribute significantly to
the integrated \Muv.  Nevertheless it is not unreasonable to assume
that if ionizing stars are present, they and stars down to
$\sim$20~\Msun\ formed at the same epoch are what dominates our
F220W observations.

Other factors affecting our F220W observations examined in the
appendix include redleak and nebular continuum contamination.  To
summarize, the F220W filter does not suffer from significant redleak,
especially when the source spectrum is young enough to ionize the ISM.
Nebular continuum emission may contribute up to $\sim 30$\%\ of the
F220W flux.  Dust can modulate the surface brightness distribution by
absorption and by scattering UV photons into and out of the line of
sight.

Figures~{1} and {4} show that the starburst
morphology in the UV is irregular.  The diffuse light distribution has
misshapen, non-elliptical isophotes.  The surface brightness profiles
(Fig.~{5}) are not well characterized by either
exponential, (except, perhaps the smooth component of
NGC~3690-Ab), or $R^{0.25}$ profiles.  Instead, like the isophotes,
the $\mu$ profiles are irregular (although one should bare in mind the
small field of view).

Embedded in the diffuse light are numerous compact sources.  These are
clearly evident in Fig.~{1}.  Those found in the course of the
smooth-clumpy separation are indicated in Fig.~{4}.  Here the
nomenclature uses the galaxy's name as the prefix, and the ranking of
the source in $M_{220}$ as the suffix. Table~\ref{t:crossid} presents a
cross-identification of the compact sources previously identified in the
literature.  In \S\ref{ss:sizes} we tabulate properties of some of the
sources, including the relative positions of all sources mentioned
below.

The peak brightness of a 100\Msun\ star is $M_{220} = -10.7$.  Sources
much brighter than this are not likely to be single stars. We show in
\S\ref{s:clust} that many of the sources are significantly brighter
than this and have sizes that are consistent with being star clusters.
We define the term ``super star cluster'' (SSC) to denote clusters
with $\Muv < -14$ (this corresponds to $M_V < -13$ for 10 Myr old
clusters; for other ages see the appendix).  These must contain at
least 20 high luminosity stars, and thus can not be confused with
small multiple star systems.

\subsection{The gallery}\label{ss:samp}

The following discussion of each galaxy's UV morphology and relevant
previous work is arranged by distance. East, north offsets (relative
to the coordinates in Table~\ref{t:obslog}) of objects in arcsecs are
given in $\pm$E.E,$\pm$N.N format.

{\bf NGC~5253}: This nearest galaxy in our sample is certainly the
Rosetta stone.  It is near enough to resolve into stars, and to
spatially resolve most of the clusters.  Without the high spatial
resolution of the HST it is hard to differentiate between clusters and
stars.  In the detailed optical study of Caldwell \&\ Phillips (1989)
many of the point like sources are incorrectly referred to as clusters.
It took the planetary camera (PC) images presented by Sandage et al.\
(1994) and Saha et al.\ (1995) to demonstrate that NGC~5253 is resolved
and that some of the sources are Cepheid variables.  In the optical, the
high surface brightness blue central region is where the emission line
strength is most intense (Walsh and Roy 1987; 1989).  We image only part
of this region; UV emission clearly extends beyond the edge of the
frame, particularly towards the south.  NGC~5253 illustrates the basic
UV morphology of a starburst: stars are distributed like the diffuse
light, while clusters (in this case easily distinguished by their fuzzy
appearance) are preferentially found near the center of the burst.  Of
the objects listed by Caldwell \&\ Phillips only the five nuclear knots
are in our frame.  Our identifications of them are listed in
Table~\ref{t:crossid}.  We do not see their knot 5.  Comparison with PC
images (Saha, 1994, private communication) suggests that it is an
emission line knot. Walsh \&\ Roy (1987, 1989) find an enhanced nitrogen
abundance and a Wolf-Rayet spectrum in their spectral cube centered near
NGC5253-1. Comparison with an {\em R} frame (kindly made available to us
by M.\ Lehnert) indicates that this cluster is the bluest of the nuclear
knots, as well as the brightest in the UV.  We tentatively identify it
as the Wolf-Rayet cluster, indicating its age is $t \approx 3 - 8$ Myr
(LH).  This object itself is resolving into stars.  A well known dust
lane causes the indentation in the isophotes to the SE of NGC5253-1.
{}From the low detected star density (Fig.~{4}) we infer that
it cuts through the center of NGC~5253, passing $\sim 2''$ to the south
of NGC5253-1.  It is the continuation into the center of NGC~5253 of the
\fion{O}{III} filament discussed by Graham (1981).  NGC~5253 is a member
of the Cen A group.

{\bf NGC~1705}: NGC~1705 is close enough to resolve into stars.  It is
the nearest galaxy in our sample that contains a SSC, and a
spectacular one at that: the third brightest in our sample if we
exclude those in the highly reddened galaxies NGC~3690 and NGC~7552.
It is discussed at great length by MFDC\footnote{The $D = 4.7$~Mpc
adopted by MFDC is in error due to a mistake in their code for
deriving $D$ from $V_r$.} and Meurer (1989), who show that it is the
likely power source of the spectacular galactic wind seen in NGC~1705.
They estimate its age as $t = 13$ Myr, which means it is no longer an
ionizing source, yet it accounts for nearly half of the total
$F_{220}$.  At an age of 2 Myr it would have been about two magnitudes
brighter (appendix).  O'Connell et al.\ (1994) present HST PC images of
NGC~1705 in the F555W and F785LP passbands from which we have
estimated the $R_e$ of NGC1705-1 (\S\ref{ss:sizes}).  After modeling
and removing NGC1705-1 we see that the underlying UV distribution is
still concentrated towards it, as can be seen by its surface
brightness profile in Fig.~{5}.  This suggests that the SSC
may have more extended wings than our simple model of a circularly
symmetric Gaussian, as would be expected for a cluster having a King
(1966) or power law profile. The removal of NGC1705-1 allows NGC1705-2
to clearly be seen.  This source is not seen in ground based images,
but is mentioned by O'Connell et al.\ (1994).  Protrusions in the the
outer isophotes of NGC~1705 correspond to knots B and C of Melnick
et al.\ (1985).  There are a swarm of sources at knot B, the brightest
\HII\ region in the galaxy (MFDC).  Some may be individual ionizing
stars.  The two UV frames are separated by 41 days.  Are there any
variable stars in NGC~1705?  Typical random errors in \muv\ from
photon statistics are 0.1 mag, while additional spatial variations in
the FOC performance will add another quasi-random error of 0.15 mag
(Meurer, 1995a).  Therefore the intrinsic uncertainty in $\Delta\muv$
between two frames is $\sim 0.25$ mag.  There are two sources which
differ by $\geq 0.5$ mag between the two exposures: NGC1705-25 has
$\muv = 19.7, 18.9$ in exposures a and b respectively; while
NGC1705-17 has $\muv = 18.9, 19.5$.  Neither are very convincing
variable star candidates.  In frame b, NGC1705-25 is at the edge of
the region where the FOC $512 \times 512$ format is burnt into the
detector, while all the sources in the vicinity of NGC1705-17 appear
dimmer in frame b relative to a, suggesting a large local zeropoint
difference.

{\bf IZw18}: With an oxygen abundance $O/H \approx 1/50$ of the solar
value (Dufour et al.\ 1988), IZw18 is the most metal poor galaxy known.
Kunth et al.\ (1994), using high resolution UV spectra obtained with
the GHRS on board the HST, argue that the oxygen abundance in the
neutral ISM is about 20 times lower than in the \HII.  However, this
interpretation is very controversial (Pettini \&\ Lipman 1995).
Our image contains the two bright condensations, commonly referred to
as the NW and SE knots, that are the usual targets of spectroscopic
studies.  Fainter structure discussed in detail by Dufour \&\ Hester
(1990) and Davidson et al.\ (1989) falls outside our image.  The two
knots are separated by $\sim 5.8''$ in our images. Despite being a
prototypical BCD galaxy, both the peak $\mu_{220}$ and the \Muv\ of
the brightest cluster are the faintest observed in our sample.  We
demonstrate in \S\ref{ss:red} that this is not due to dust extinction.
Compact sources are seen in both knots; about three times more in the
NW knot than SE knot.  IZw18 is resolving into stars at the detection
limit of our deep image.  A faint filament, $\sim 1.8''$ long, appears
to emanate westwards from the SE knot and then curves to the NW.

{\bf NGC~4670}: Recently Hunter et al.\ (1994b) presented a detailed
\HI\ and optical study of NGC~4670 which shows that high mass star
formation is restricted to a single central starburst.  Outside of the
starburst, the galaxy has smooth elliptical isophotes in the optical.
This is very similar to other (but lower luminosity) amorphous/BCD
galaxies such as NGC~1705 and NGC~5253.  Our image is centered on the
central kidney shaped starburst with dimensions of $880 \times 470$
pc.  It contains 10 SSCs and numerous fainter sources. Low surface
brightness extensions, $\sim 100-200$ pc wide and suggestive of a bar,
continue eastwards and westwards to the edges of the frame. A
moderately bright cluster, NGC4670-21, is contained in the westwards
extension. A faint diffuse (diameter $\sim 0.6''= 60$ pc) UV source
located at --10.0,+3.9 corresponds to a detached \HII\ region seen in
the images of Hunter et al.\ (1994b).

{\bf NGC~3310}: We image the central portion of this well studied
starburst spiral.  Its structures in the optical continuum, \Halpha,
infrared, and radio continuum (e.g.\ Balick \&\ Heckman 1981; Telesco
\&\ Gatley 1984) are reasonably well correlated with the UV.  They
include a central ring with $r \approx 7'' = 610$ pc, a mildly active
nucleus, and the ``Jumbo'' \HII\ region of Balick \&\ Heckman.  The
latter is known to contain Wolf-Rayet stars (Pastoriza et al.\ 1993)
and is located where NGC~3310's southern arm attaches to the central
region.  These features are seen in our image, although the jumbo
\HII\ region is not completely in the field of view.  The SE rim of
the ring clearly has a higher surface brightness than the ill defined
NW rim which is composed of two faint strands separated by $\sim
5''$. Overall the morphology is suggestive of tightly wound spiral
arms instead of a ring (also noted by van der Kruit \&\ de Bruyn,
1976).  The outer NW strand defines the stellar ring of Balick and
Heckman; the nucleus is near the center of this ring.  However, the
dynamical center is near the center of the ring defined using the
inner NW strand (van der Kruit 1976; Grothues \&\ Schmidt-Kaler,
1991).  Several dust lanes cut through the ring, most noticeably at
+4.1,--4.3 and +1.9,--6.2.  The nucleus is resolved: $\Wfo
\approx 1.4'' = 120$ pc and contains a peculiar right angle structure
(dimensions $39 \times 46$ pc; width $\lapeq 9$ pc).  The cluster
distribution is like that in the AGN galaxies NGC~1097 and NGC~6951
(Barth et al., 1995): clusters are found along the starburst ring
(including the Jumbo \HII\ region which contains at least six) but not
interior to it.

{\bf NGC~7552}: We image the center of this nearly face-on spiral that
shows ring and bar like structures on several scales (Feinstein et al.,
1990; Forbes et al.\ 1994a,b; hereafter F94a, F94b).  The most relevant
to us is the $9'' \times 7'' = 960 \times 670$ pc nuclear ring at the
inner Lindblad resonance.  This is best seen at radio wavelengths and
in \Brg\ emission, and optical and infrared color maps, but is
difficult to detect in either optical or IR broad band images
(F94a,b).  Interior to this is a nuclear bar which shows up in $1-0$
S(1) H$_2$ emission (F94b).  The nucleus itself is weak or absent at
all wavelengths leading F94b to conclude that it is dormant. In the UV
there are two emitting regions corresponding to two of the hot spots
along the nuclear ring. The brightest is nearly circularly symmetric
at moderate surface brightness levels, with an extension to the south
at lower levels.  It corresponds to the northern part of the radio
knot A, the brightest \Halpha\ hotspot, and the ``blue break'' in the
optical images of F94a.  In the UV its center divides into two bright
clouds 70 pc and 140 pc in diameter, perhaps separated by a dust
lane. The larger of the two has four of the brightest SSCs,
NGC7552-1,3,4,5, scattered around its periphery and the smaller has
NGC7552-2 at its center.  There are numerous fainter clusters
throughout the halo of this region, in particular a spray of sources
1\as\ -- 3\as\ S of the brightest source.  The sources closest to the
expected position of the nucleus (at --2.8,--1.6) are NGC7552-13,25
and have a relatively inconspicuous $\Muv \approx -14$.  The other UV
bright region (--5.0,--2.1) corresponds to radio knot C of F94a.  It
contains four clusters.  Much of the UV morphology of NGC~7552 is
probably dictated by location of holes in the distribution of dust.
We know from the radio and \Brg\ rings that a starburst ring is
present (F94a,b).  But at the two brightest \Brg\ hotspots, (B and D
in the notation of F94a), where the ionizing emission should be
strongest, there is little UV emission.  The hotspots seen in the UV
are those with the lowest extinction measurements in F94b.

{\bf Tol1924-416}: Bergvall's (1985) detailed optical study of this
galaxy shows it to be dominated by a central starburst $\sim 1.8$ Kpc
across, which is embedded in a low surface brightness host.  The
central complex has two peaks separated by $\sim 540$ pc on an EW
line.  Iye et al.\ (1987) discuss the emission line kinematics in this
region. Bergvall notes that there may be some broad band variability
associated with Tol1924-416, but this has not been confirmed. The
central complex has a footprint morphology ($1.8 \times 1.0$ Kpc).
The eastern peak of Bergvall clearly corresponds to Tol1924-416-1, one
of the brightest SSCs in our sample.  There are several diffuse
sources associated with the western peak.  A LSB knotty plume, 890 pc
long, is seen extending from the heel to $\phi \approx 210\dg$.  There
are at least three isolated sources.  The brightest, Tol1924-416-2, is
double (companion Tol1924-416-12) and clearly seen in ground based
images.  It may have faint \Halpha\ emission associated with it
(Bergvall, 1985) suggesting that it is in the starburst and not a
foreground star.

{\bf NGC~3690}: The merging system Arp~299, consisting of NGC~3690 and
IC~694, is very well studied (e.g.\ Gehrz et al.\ 1983; Friedman et al.\
1987; Joy et al.\ 1989; AHM; Wynn-Williams et al.\ 1991, Mazarella and
Boroson, 1993).  Our image centered on NGC~3690 is one of the more
spectacular images in our sample.  It shows three UV bright clouds.
The largest we refer to as ``BC'' because it contains the sources B
and C of Gehrz et al.\ (1983).  The smaller clouds are ``Ab'' and
``Aa'' following the nomenclature of Mazarella \&\ Boroson (1993).
The three clouds have $\sim 60$ SSCs scattered amongst them (see
Table~\ref{t:crossid} for cross identification with previous work).
BC, is shaped like a foot with dimensions of $2.4 \times 1.5$ Kpc.
The bright \Halpha\ emission regions correspond to the double SSC
NGC3690-5,8 in the ``heel'' and the numerous SSCs in the ``toes'' of
this structure such as NGC3690-6,7,9,10 (Wynn-Williams et al.\
1991). Stanford \&\ Wood (1989) measure an \HI\ column density of
$N_{\HI} = 3.1 \times 10^{21}\,{\rm cm}^{-2}$ towards BC.  Using the
conversion factor of Burstein \&\ Heiles (1978) this corresponds to
$E(B-V) = 0.62$, in excellent agreement with \ebvbd\
(Table~\ref{t:redex}) but not \ebvuv\ derived below.  The two detached
clouds Ab (--7.4,--4.6) and Aa (--13.3,--3.7) are each more luminous
($M_{220} = -19.9, -19.6$ respectively) than the dwarf galaxies in our
sample: NGC~1705, IZw18, and NGC~5253.  However, the extinction
correction is large, and long-slit spectroscopy indicates that the
reddening is not uniform, and thus \Muv\ may be underestimated (Gehrz
et al.\ 1983; Friedman et al.\ 1987; Mazarella \&\ Boroson 1993).  Ab
and Aa are both highly concentrated towards two of the most luminous
SSCs (NGC3690-4,3 respectively), and also contain other clusters.
Inadvertently, Aa was occulted by the F/96 thin finger.  There were
two recent SNe in NGC~3690: SN1992bu (van Buren et al.\ 1994), and
SN1993G (Treffers et al.\ 1993; Forti 1993).  The position of
the latter is outside our field of view, while there is nothing
apparent within 2\as\ of the published position ($\sim +5.6,-8.0$ in
our coordinate system) of the former.  This is not surprising since
SN1992bu exploded over a year prior to our observation.

{\bf NGC~3991}: This outlying member of the NGC~3991/4/5 group
(Garcia, 1993) is composed of two clumpy star forming complexes,
aligned along either side of a faint nucleus (Keel et al.\ 1985).  The
northern complex is clearly the brighter of the two, and is all that is
contained in our image.  It has a bent peanut morphology (similar to
IZw18) with a bright (northern most) and faint component separated by
1.9 Kpc.  The faint component is partially occulted by the F/96 wide
finger.  NGC~3991 contains about two dozen clusters. A faint thin
($\approx 0.1'' = 22$ pc wide) filament 830 pc long apparently is
directed towards (or from) the elongated source NGC3991-8.

\section{Integrated properties}\label{s:prop}
\subsection{Spectral slope and UV extinction.}\label{ss:red}

One of the biggest obstacles to interpreting UV observations is dust.
It scatters and absorbs UV radiation much more efficiently than
optical/IR radiation.  The scattering can be both into and out of the
line of sight.  The net amount of light removed from the line of sight
as a function of wavelength, the extinction law, depends critically on
both the geometry of the dust distribution and the dust composition.
Dust sufficiently close to a starburst will reradiate the light it
absorbs in the far infrared.  Thus the observed ratio of far infrared to
UV emission, the far infrared excess, is a diagnostic of the
redistribution of the spectral energy by warm dust.  The spectral slope
or color is a strong diagnostic of the dust geometry.  If the dust
distribution is inhomogeneous, or the dust is mixed in with the stars,
the transmitted UV spectrum will be weighted towards the least extincted
lines of sight; consequently the spectrum will be bluer than if the dust
were in a foreground screen (Witt et al.\ 1992).

Figure~{6} shows ${\rm IRX} \equiv \log(\FIR/F_{220})$ plotted
against the UV spectral slope, $\beta$ ($f_\lambda \propto
\lambda^\beta$) for starburst galaxies.  Here,
\FIR\ is derived from 60\micron\ and 100\micron\ IRAS observations
following the the definition of Helou et al.\ (1988).  $F_{220}$ is the
F220W UV flux ($\lambda f_\lambda$, as defined in the appendix),
corrected for Galactic extinction (see Table~\ref{t:redex}), but not
the internal extinction associated with the starburst.  The data from
our sample are supplemented with the remainder of galaxies in the K93
catalog which have IRAS fluxes.  The fluxes for our program galaxies
are reported in Table~\ref{t:flux}.

The strong correlation in Fig.~{6} is in the sense that the
redder $\beta$ is, the higher \FIR\ is relative to $F_{220}$.  This is
exactly what is expected for reddening by a foreground screen of dust,
as we show below.  The horizontal bars in Fig.~{6} show the
range of intrinsic $\beta$ expected for ionizing stellar populations,
$\beta_0 = -2.5 \pm 0.2$ (appendix).  This is the {\em a priori\/}
expected $\beta_0$ under the ionizing population assumption, and
effectively the lower limit to the observed values of $\beta$; the
majority of the galaxies have redder $\beta$, as would be expected if
they were reddened by intervening dust.  Calzetti et al.\ (1995) find a
similar correlation between $\beta$ and $\log(L_{\rm IR}/L_{\rm
B})$. Further evidence that the UV slopes are reddened is presented in
Fig.~{7} (and Fig.~12 of C94) which shows that $\beta$
correlates with the reddening derived from the Balmer decrement,
\ebvbd\ (tabulated in Table~\ref{t:redex}).  The two bluest galaxies
in our sample, NGC~1705, and IZw18 have $\beta = -2.5$, so they are
virtually unreddened.  Both galaxies also have a very low
\FIR, suggesting a low dust content.  IZw18 is undetected by IRAS,
while MFDC estimate $\ebv \leq 0.01$ for NGC~1705 from \FIR, assuming
a foreground screen geometry for the dust.

Figure~{6} provides constraints on the dust distribution and
extinction law applicable to starburst galaxies.  The dust has to be in
the environment of the starburst (i.e.\ not in our Galaxy or the
intergalactic medium) otherwise the dust would not be heated and IRX
would be low for all sources. The large range of observed $\beta$ rules
out a grey extinction law, or extinction only by UV opaque clouds of
dust (or ``bricks''), since in those cases we expect all galaxies to
have $\beta = \beta_0 \approx -2.5$.  It is also unlikely that the dust
is evenly distributed and purely internal to the UV emitting region.  In
that case the UV emission and spectral slope will be dominated by stars
within one optical depth of the ``front'' of the starburst and $\beta$
will asymptote to a constant value for high dust content.  C94 model
this geometry and find that the maximum increase in $\beta$ is 0.6, and
1.05 for Milky Way and LMC extinction laws, respectively.  The
starbursts with $\beta > -1.2$ can not be explained with this geometry,
nor can the good correlation between $\beta$ and IRX.

Figure~{6} shows the expected correlations for models
where the dust is in a uniform foreground screen and the extincted
radiation from $\lambda 912$\AA\ to $\lambda 8000$\AA\ is reradiated
in the far infrared.  We assume that 71\%\ of this flux will be
intercepted by the \FIR\ ``passband''.  This is appropriate for dust
at $T \approx 40-80\,$K and a dust emissivity $\propto \nu$ (Helou
et al.\ 1988).  The source spectrum is assumed to be the total
stellar plus nebular spectrum of a 10 Myr old constant star formation
rate population (solar metallicity, Salpeter IMF) from the models of
LH, and has $\beta_0 = -2.5$.  Four reddening laws are considered: the
Galactic reddening law of Seaton (1979), the LMC reddening law of
Howarth (1983), and the starburst extinction laws of Kinney et al.\
(1994; hereafter K94) and C94.  The appendix parameterizes the F220W
extinction, $A_{220}$, and reddening of $\beta$, as a function of
\ebv\ for these reddening laws.  Here we assume that the UV
``aperture'' (i.e.\ the extraction aperture of K93, or the FOC frame)
recovers most of $F_{220}$, so that dust scatters equal amounts of
flux into and out of the integrated line of sight.  In other words,
the absorption law is the extinction law.  This is consistent with how
the the C94 and K94 extinction laws were derived, although not how the
Galactic and LMC laws were derived.  If a source spectrum with
different $\beta_0$ is adopted, the primary effect on
Fig.~{6} is to shift the theoretical curves horizontally,
with very little change in the shape of the curves, at least for
ionizing populations.  The LMC, C94 and K94 extinction models follow
the observed correlation very well considering the simplicity of the
models.

The importance of Fig.~{6} is that it demonstrates that at
least some of the UV to IR redistribution of light is due to
foreground dust in close proximity to the starburst.  How close must
the dust be?  We calculate a rough estimate as follows.  The models of
D\'esert et al.\ (1990) predict a tight relationship between the ratio
of 60\micron\ and 100\micron\ IRAS fluxes $f_\nu(60)/f_\nu(100)$ and
the UV energy density, $U$:
\[ \log(f_\nu(60)/f_\nu(100)) = -0.48 + 0.30 \log U, \]
where $U$ is given in eV cm$^{-3}$, and the above relationship is good
for $\log U < 3$, and calibrated for O5 and B3 stars.  For an
unattenuated point source $U = L_{\rm UV}/ (4\pi R^2 c)$.  The IUE
sample of starbursts has $\langle \log(f_{60}/f_{100}) \rangle =
-0.17$.  Adopting this mean, and assuming that the intrinsic spectrum
is a power law with $\beta_0 = -2.5$ and $L_{\rm UV}$ is the
integrated flux from 912\AA\ to 3000\AA, then in terms of the F220W
luminosity, the radius of the dust shell is
\[ R_{\rm Dust} \approx 190 \sqrt{\frac{L_{220}}{10^{42}
{\rm erg\, s^{-1}}}}\, {\rm pc}. \]
For the luminosity range of the FOC sample $0.3 < L_{220}/(10^{42}) \,
{\rm erg\, s^{-1}} < 90$ we have $0.1 < R_{\rm dust}/{\rm (Kpc)} < 1.8$.
We find a median $R_{\rm Dust}/R_e$ $= 2.5$.  Thus in this model most of
the dust is significantly beyond the effective radius, but still
fairly close to the starburst.

Up until at least a decade ago, a foreground screen geometry for the
dust was the standard implicit assumption of most astronomers.
However, this geometry has been gradually rejected as being too
simplistic.  Weedman (1988, 1991) noted that UV selected galaxies had
far IR properties similar to those selected for their far IR strength.
For galaxies to be both UV and IR bright suggests that we are
preferentially seeing galaxies through unobstructed lines of sight.
Following up on this notion Witt et al.\ (1992) explored radiative
transfer through a variety of geometries more complicated than a
simple foreground screen.  They note that ``the screen geometry leads
to a phenomenon that is simultaneously both widely believed and
implausible: reddening of broad band color is positively correlated
with increasing dust extinction.''  Figure~{6} shows exactly
that phenomenon.  How can this be?

First of all, a foreground screen geometry may not be so implausible
after-all.  The central region of a starburst is not likely to be a
hospitable environment for dust, except in dense molecular clouds
which will have a low volume filling factor.  The Galactic winds
frequently observed in starbursts will sweep out any diffuse ISM from
the immediate environs of the starburst on the timescale of a few Myr
(Heckman et al.\ 1990).  A cavity around the starburst will form filled
with a hot plasma of thermalized SNR having $T \sim 10^7 - 10^8$ K.
Within the cavity there may be some molecular clouds compressed to
high density and low volume filling factor by the hot plasma.  As
these evaporate they may release dust into the cavity.  However,
assuming a typical particle density of $n = 0.1\,{\rm cm}^{-3}$ in the
cavity the timescale for destruction by sputtering is $\sim 5$ Myr for
grains having a relatively large size of 0.25\micron\ (Draine and
Salpeter, 1979).  Thus dust will be destroyed on a timescale
equivalent to the time required to transport it from the core and
somewhat less than the star formation timescale (see below).  Much of
the dust may be in the shell swept up by the Galactic wind.  This
shell would make an ideal foreground screen being roughly uniform
until it fragments.  In addition, part of the foreground dust may be
in relatively undisturbed ISM remaining outside the starburst.  We
would expect this ISM to be from the portion of the starburst's
nascent cloud not involved with star formation, and perhaps ``fresh''
ISM flowing in to ``feed'' the starburst.

Second of all, it is clear that a homogeneous foreground screen is not
the whole story.  There is plenty of evidence for some of the dust to
be clumped, or patchy.  Figure~{7} shows the K94 reddening
vector in the $\beta$, \ebvbd\ plane assuming an intrinsic $\beta_0 =
-2.5$ and assuming $E(B-V)_{\rm UV} = E(B-V)_{\rm BD}$, as we would
expect for a uniform screen dust geometry.  The correlation is
significantly shallower than this vector. This is a well known
phenomena: the Balmer decrement overpredicts the UV extinction
(Fanelli et al.\ 1988; C94).  It indicates that in starburst galaxies
the \HII\ has a different distribution relative to the dust than the
UV.  Thus a uniform foreground screen geometry is not totally
adequate.  For our sample the most discrepant points from the K94 line
in Figs.~{6},{7} are NGC~3690 and NGC~7552. From
NGC~7552 we are seeing only a fraction of the UV flux of the starburst
ring of F94a, F94b as explained in \S\ref{ss:samp}.  In this case the
UV morphology is dictated by the location of holes in the dust
distribution.  It is then no surprise that $\beta$ is steeper for the
FOS spectrum, centered on the UV surface brightness peak, probably
where the dust coverage is thinnest, than the IUE spectrum which
includes the more extincted ``edges'' of the hole
(Table~\ref{t:spprop})\footnote{So in this case the foreground screen
is not the whole story but instead the {\em hole\/} story!}. In
NGC~3690, Gehrz et al.\ (1983) detect a strong 11.4\micron\ silicate
feature indicating that much of the mid-infrared flux is due to
features extincted by 5 to 14 magnitudes in the visual.  The optical
and UV data suggest extinctions of only 2 mag.  Indeed, NGC~3690 is
one of the most discrepant points in Fig.~{6}, in the sense
we would expect if some of the UV sources are completely obscured.
Dust lanes seen in our images of NGC~5253, NGC~3310 and NGC7552,
further demonstrate that the dust distribution is not uniform in
starbursts.

The dust distribution is thus not easily categorized.  Although
patchiness is directly observed in many cases, a significant fraction of
the dust must be in a foreground screen.  This is even true where the
evidence for patchiness is greatest, NGC7552 and NGC~3690, otherwise
they would not appear reddened.  Since a foreground screen geometry
adequately models the IRX - $\beta$ correlation over large apertures, we
can use this model to estimate the {\em total} UV fluxes from the
measured fluxes and $\beta$.  This model is not sufficient to estimate
either the total dust content or the local extinction. Nevertheless, out
of a lack of a better solution, at present, we will adopt a uniform
extinction for each galaxy.

For the remainder of the paper we adopt the starburst UV extinction
law of K94 to derive the internal extinction.  It is very similar to
the LMC extinction law of Howarth (1983).  This is appealing because
the Howarth law does not in fact work for all of the LMC, but instead
is applicable only to a $\sim$500 pc region surrounding 30 Dor
(Fitzpatrick, 1985), the nearest starburst to us.   The main
difference between the Howarth and K94 laws is the lack of the
$\lambda 2175$\AA\ bump in the K94 law.  Figure~{6} shows that
this law underpredicts IRX for a given $\beta \gapeq
-1$.  This is consistent with additional patchiness in the dust
distribution.

Here, the internal color excess is taken to be $\ebvuv = (\beta -
\beta_0)/8.067$ (appendix), with $\beta_0 = -2.5$, as expected for
ionizing populations.  As discussed in the appendix, a wide range of
star formation histories, from short bursts less than 10 Myr old to
continuous star forming regions 100 Myr and older will produce a
spectrum with very similar $\beta_0$.  The range in $\beta_0$ of $\pm
0.2$ for ionizing populations amounts to an uncertainty in $\ebvuv$ of
$\pm 0.025$ mag, and in $A_{220}$ of $\pm 0.21$ mag.  The adopted
internal color excesses \ebvuv\ and $E(\beta) = \beta - \beta_0$ are
given in Table~\ref{t:redex} along with the consequent total (Galactic
+ internal) extinction, $A_{\rm 220,T}$.  For these calculations we
use the $\beta$ values corrected for Galactic extinction given in
Table \ref{t:spprop}.  These are the IUE values except for NGC~3690,
where the IUE aperture is probably not centered on NGC~3690
(\S\ref{ss:reduc}). For it we use the FOS $\beta$.

\subsection{Intrinsic properties.}\label{ss:glob}

Table \ref{t:intrinsic} presents some integrated intrinsic (i.e.\
extinction corrected as discussed above) properties of the sample.
The radii $R_e$ and $R_{\rm max}$ are mean radii derived from $a_e$
and $a_{\rm max}$ using
\[ R = \sqrt{ab}. \]
The UV half light radii of the starbursts are in the range $100 \lapeq
R_e \lapeq 700$ pc with the exception of NGC~1705.  This is because
NGC1705-1 contains nearly half the total flux of the galaxy.  Since
its size is not apparent from the UV observations, we report two cases:
(1) it is assumed to be a point source (2) NGC~1705 in the absence of
NGC1705-1. The size range of our sample is somewhat smaller than that
of FIRGs,  the largest of which have $R_e \sim 2$ Kpc (AHM).  This may
be in part because we resolve one of the largest starbursts in our
sample, NGC~3690, into subcomponents.

The next two columns in Table~\ref{t:intrinsic} report UV luminosity
$L_{220} = 4\pi D^2 F_{220,0}$, and $\mu_{e,0}$, the mean extinction
corrected surface brightness within $R_e$.  The remaining columns
convert these to quantities involving stellar mass.  To do this we
employ the population models presented in the appendix which use an
IMF with a Salpeter (1955) slope ($\alpha = 2.35$).  We primarily
calculate the mass in massive stars, $M_{M*}$, defined to be those in
the mass range $m_l = 5\,\Msun$ to $m_u = 100\,\Msun$, since our
observations are not sensitive to stars with $m_* < 5\,\Msun$.
Detailed modeling of the properties of the M82 starburst indicate that
its IMF may be deficient in low mass stars (Rieke et al.\ 1980; 1993;
McLeod et al.\ 1993), so our conservative approach is warranted. If the
actual lower limit to the IMF is $m_l = 1\,\Msun$ or 0.1~\Msun, then
$M_{M*}$ underestimates the total stellar mass by a factor of 2.16 or
5.52 respectively.  For ionizing populations neither $L_{220}$ nor
$M_{M*}$ depend strongly on $m_u$ for $m_u > 60$~\Msun.  For example
in a constant star formation rate population 10~Myr old, with a
Salpeter IMF, and the above $m_u$ and $m_l$, only 11\%\ of the mass of
the IMF and 18\%\ of $L_{220}$ comes from stars with $m_* > 60$~\Msun.

The conversion from $L_{220}$ to mass also depends on the star
formation history of the population.  The most important constraint we
can place on the star formation history comes from the size of the
bursts.  Simultaneous star formation in an extended region is unlikely
to occur over timescales shorter than the crossing time.  By this we
mean the time it takes for some disturbance that may effect the star
formation rate to pass through the ISM.  Taking the typical sizes and
gas velocities observed for starbursts, the typical crossing times are
on the order of 10 Myr; i.e.\ diameter of 500 pc, velocity dispersion
of $\sim 50$~\kms\ (MFDC, Iye et al., 1987, Marlowe et al., 1995).  This
provides a rough lower limit to the burst duration.  It is too long to
be an instantaneous starburst (ISB), so we adopt a constant star
formation (CSF) rate history for the global properties of the
starbursts.  As noted below, the individual clusters are small enough
that they can form over a much shorter timescale, and are likely to be
formed in true short duration bursts.

For deriving masses we adopt $M_{M*}/L_{220} = 0.0017\, (\Msun/ L_{\rm
Bol})_\odot$ which is the average mass to light ratio for CSF models
with ages $1 < t \leq 100$ Myr and our adopted mass range.
$M_{M*}/L_{220}$ covers a range of a factor of 13 for the full range
in star formation histories of ionizing populations.  Note $M_{M*}$ is
the total {\em initial\/} mass in massive stars.  The instantaneous
mass in stars is somewhat less due to stellar winds and SNe returning
mass to the ISM (see LH for detailed calculations).

The conversion from $L_{220}$ to the massive star formation rate,
MSFR, has only a weak time dependence for a CSF history.  We adopt
\[ {\rm MSFR} = \frac{L_{220}}{2.70 \times 10^{43}\, {\rm erg\,
s^{-1}}} \Msun\, {\rm
yr}^{-1}. \] The conversion factor is consistent with the above
$M_{M*}/L_{220}$ ratio for an age of 12 Myr (conveniently close to the
crossing time estimated above).  Similarly the effective massive star
formation activity, \msfae, is the average MSFR per unit area within
$R_e$;
\[\msfae = \frac{\rm MSFR}{2\pi R_e^2}, \]
Again we stress that MSFR and \msfae\ only measure the contribution
of massive stars. A suitable correction factor, such as those given
above, must be applied to get the total star formation rate.  For
comparison column 9 of Table 9 gives SFA$_e$, the star formation rate
per area assuming the IMF extends down to $m_l = 0.1$~\Msun.

Figure~{8} plots some of the integrated properties of the
sample.  Panel (a) plots \msfae\ against $\log(L_{220})$, and panel (b)
plots \msfae\ against $\log(R_e)$.  Most of our sample has $\langle
\log(\msfae) \rangle = -0.16$ (or $\msfae = 0.7 \Msun\,{\rm Kpc^{-2}
yr^{-1}}$) with a standard deviation of 0.16. This corresponds to
$\mu_{e,0} = 15.9 \pm 0.4$ mag arcscec$^{-2}$.  The exception are
three outliers, IZw18, NGC~1705 (total), and NGC~7552.  The high
surface brightness of NGC~1705 is due to NGC1705-1.  If it is excluded
NGC~1705's \msfae\ is normal relative to the other starburst galaxies.
NGC~7552 has the highest \msfae\ in the sample, significantly outside
(by $6\sigma$) the range of the majority of starbursts.  The high
surface brightness of NGC~7552 is immediately apparent in the FOC
image (despite its high $A_{220}$) and causes this image to have the
highest degree of nonlinearity.  However, because of the large
extinction and non-linearity corrections, its quoted value for \msfae\
should be considered tentative.  The low \msfae\ of IZw18 ($5\sigma$
lower in $\log(\msfae)$ than the majority of the sample) can not be
explained by mitigating circumstances.  Including NGC~7552 and IZw18
in the average does not change $\langle \log(\msfae)
\rangle$ significantly but increases the dispersion to 0.4.  Panel (c)
of Fig.~{8} shows a strong correlation between
$\log(L_{220})$ and $\log(R_e)$.  The dotted line in panel(c) shows
the expected slope for constant constant surface brightness.

Panels a-c of Fig.~{8} illustrate that most starbursts have
roughly the same UV surface brightness and consequently that $L_{220}$
is mainly governed by the size of the star forming region.  AHM find
a similar result for FIRGs selected to have warm dust temperatures;
specifically that $L_{\rm IR} \propto R_e^{1.7 \pm 0.2}$ where here, $R_e$
is the \Halpha\ half light radius. This is close to the $R_e^2$
correlation expected for constant surface brightness.  The UV and IR
results are important because they suggest a negative feedback
mechanism is limiting the \msfae.  Since we selected a starburst
sample, we may be selecting galaxies with a high \msfae.  Thus the
mean \msfae\ we find for starbursts may represent an upper limit for
the larger encompassing set of star forming galaxies.  In comparison
normal disk galaxies have a mean \Halpha\ surface brightness $\langle
\Sigma_{\rm H\alpha}\rangle = 3.6 \times 10^{39} {\rm erg\, s^{-1}\,
Kpc^{-2}}$ (Kennicutt, 1989) which corresponds to ${\rm MSFA}_e =
3.8\times 10^{-3}\, {\rm \Msun\, Kpc^{-2}\, yr^{-1}}$ (for a 10 Myr
CSF model; LH and appendix). On average our starburst sample has star
formation activity about 200 times more intense than normal disk
galaxies, thus earning their ``starburst'' moniker.  In addition,
Lehnert (1992) finds that the range in $L_{IR}/R_e^2$ for the warm
starbursts (i.e.\ those selected by AHM), is much smaller than for
FIRGs selected with no color constraints, and that the warm FIRGs have
the highest $L_{IR}/R_e^2$.

In panel (d) of Fig.~{8} the extinction corrected UV and
\Halpha\ fluxes are compared.  The fluxes we use are listed in
Table~\ref{t:flux}.  The \Halpha\ fluxes were taken from the
literature using the largest appropriate aperture.  For NGC~3991 and
Tol1924-416 the available \Halpha\ data employed quite small
apertures.  Therefore we adopt an $F_{220,0}$ through an aperture with
matching area for this comparison.  Note that the extinction
correction adopted to determine \Foha\ is that derived from the Balmer
decrement, \ebvbd, not \ebvuv.  Since both $F_{220,0}$ and \Foha\
measure the hot star content, they are indicators of the MSFR.  The
\Halpha\ flux is an indicator of the ionizing stellar flux, while
$F_{220,0}$ is sensitive to somewhat less massive stars, and therefore
longer star formation durations.  Also shown in Fig.~{8}d are
the correlations expected from the LH models for constant star
formation rate populations with ages 1, 10, and 100 Myr, and a
Salpeter IMF with $m_u = 100 \Msun$.  The UV and \Halpha\ fluxes have
the one to one correlation expected, but have, on average a higher
$F_{220,0}$ for a given \Foha\ than the LH models by a factor of
several.  There are several possible explanations for this
discrepancy.  The starburst galaxies may have relatively more low mass
stars than the LH model either because (1) $m_U$ is lower than
100~\Msun\ (cf.\ Doyon et al.\ 1992), (2) the IMF slope may be steeper
than a Salpeter IMF, (3) the star formation has lasted longer than 100
Myr, or (4) the MSFR is declining with time.  Other explanations for
the discrepancy include (5) Some of the ionizing flux is absorbed by
dust grains (c.f.\ Panagia, 1977; Fig.~4), (6) the nebula is density
bounded, or (7) the total \Halpha\ flux is not intercepted by the
apertures.

\section{Star clusters}\label{s:clust}

Figure~{9} shows histograms of the magnitudes of compact
sources found using DAOFIND, during the course of the smooth-clumpy
structure separation discussed in \S\ref{ss:reduc}.  Only sources with
aperture photometry having $S/N > 5$ are shown here.

The range of $M_{220}$ shown in the histograms $\sim -9$ to $\sim
-18$, shows that while some of the faintest sources in the nearest
galaxies could be stars, the brightest certainly are not and must be
clusters of some sort.  Since an individual star cluster is expected
to form over a short time ($dt < 0.1$ Myr; Larson, 1988), it can
safely be assumed to be an instantaneous starburst.  Assuming a
Salpeter IMF slope, a lower limit to their massive star content can be
found by assuming that they are at their peak UV luminosity; then
$M_{M*}/L_{220} = 5.8 \times 10^{-4}\,\Msun/L_{\rm \odot,Bol}$. The
minimum $M_{M*}$ covers a range of $6\times 10^3 - 2.4 \times 10^{5}$
\Msun, for \Muv\ in the range --14 to --18 mag covered by the SSCs.
If the IMF extends down to 0.1\Msun, then the range is from $3.3\times
10^4\, \Msun$ to $1.3 \times 10^6\, \Msun$.  This is the same mass
range found in Galactic globular clusters.

\subsection{Location}

The clusters are preferentially found where the underlying surface
brightness is highest.  One indication of this phenomenon is shown in
Fig.~{10} which plots the fraction of total UV light arising
from clumpy structure $f_{\rm clumpy}$ as a function of \msfae.  These
quantities are weakly correlated, having a correlation coefficient $R
= 0.45$ (we have excluded the total NGC~1705 measurement which is
dominated by NGC1705-1).  The correlation is in the sense that as
\msfae\ increases a greater fraction of the luminosity is produced in
compact sources.  One must be careful to not over interpret this
result.  $f_{\rm clumpy}$ is the ratio of the light in compact sources
to the total light.  There is no \Muv\ cutoff to what is called a
compact source, so in the nearest galaxies, especially NGC~5253, there
is a significant contamination by individual stars.  Also since most
of the sample occupies a narrow range in \msfae\ (\S\ref{ss:glob}), the
correlation is largely driven by the two outlying points, IZw18 and
NGC~7552.

It is more clear that the brightest clusters are {\em locally\/}
correlated with the highest UV surface brightnesses.  This is shown in
Fig.~{11} which plots \Muv\ against the local underlying
surface brightness for each galaxy, measured from the smooth component
image.  The distribution of points in each panel is limited at the
bottom by the detection limit, and on the right by the minimum
$\mu_{220,0}$ of the frame.  But there are no detection limits placed on
the upper left of each panel, that is for bright objects on top of a
weak underlying background.  That this region is underpopulated in all
panels can not be due to selection effects.  Figure~{12}a shows
the combined \Muv\ versus $\mu_{220,0}$ plot for the SSCs ($\Muv < -14$)
in NGC~1705, NGC~4670, NGC~3310, Tol~1924-416, and NGC~3991.  The other
galaxies were excluded because either (1) they didn't have luminous
enough clusters (IZw18, NGC~5253), or (2) they have high differential
reddening and/or linearity corrections (NGC~7552,
NGC~3690). Figure~{12}b is a histogram of the number of these
clusters as a function of $\mu_{220,0}$.  It is sharply peaked at a
MSFA$\approx 0.9\, \Msun\, {\rm Kpc^{-2}\, yr^{-1}}$.  The width of the
distribution is partially due to the slightly different limiting surface
brightnesses of each frame.  Panels c and d of this figure are the same
as a and b except the abscissa is $\mu_{220} - \mu_{\rm 220,min}$.  We
see that 29\%, 76\%, and 93\%, of the SSCs are located within 0.5,
1.0, and 1.5 mag arcsec$^{-2}$ of $\mu_{\rm 220,min}$, the peak surface
brightness.  In comparison 11\%, 25\%, and 49\%, on average, of the
smooth component light is located within these isophotes.

\subsection{Cluster sizes}\label{ss:sizes}

We measured the half light radius, $R_e$ of 85 compact sources.
Selection for measurement was as follows.  For the nearest two
galaxies (NGC~5253, NGC~1705), the five brightest uncrowded objects
were measured as well as the noticeably diffuse sources.  Two or three
point source candidates were also measured to test the limits of our
technique.  For the remaining galaxies, selection was by brightness.
Our target was to measure all uncrowded sources with $M_{220} < -14$
(i.e.\ the SSCs).  This was not feasible for NGC~3690 and NGC~7552
because of their high reddening correction.

Our method is to compare the objects' radial surface brightness
profiles, $S_o(r)$ (in counts per pixel) to profiles of model clusters
that have been convolved with the PSF.  Examples of fits are shown in
Fig.~{13}.  The object profiles were extracted from the
linearity corrected images\footnote{For NGC~1705, the frames with
NGC1705-1 subtracted were used.} in one pixel wide circular annuli.
After some experimenting we decided to discard the central annulus
($r$ ranging from 0 to 1 pixels) from the fit because of the onset of
non-linearity or saturation in many of the brighter sources.  The
worst case of this is NGC1705-1 which is so saturated that we could
not fit our UV data with this technique.  Tol1924-416-1 (shown in
Fig.~{13}) is the second worse case, it has a flat $S_o(r)$
out to $r=2.5$ pixels, where we begin our fit.  In most cases the
profile was fit out to $r=8.5-10.5$ pixels; the exact value depended
upon crowding.  In one case, NGC5253-12 (also shown in
Fig.~{13}), the fit was extended to $r=15.5$ pixels because
of the diffuseness of the cluster.  Sources with neighbors of at least
comparable brightness within $r = 12$ pixels were not fitted.
However, apparently elongated sources or those with faint extensions
were not excluded.  These may represent sources with faint companions.

The clusters were modeled by circularly symmetric Gaussian profiles.
For this model $R_e = 0.5\Wfo$.  Two sets of models were made by
convolving the profiles with the two empirical PSFs. A $\chi^2$
minimization technique was used to fit each $S_o(r)$ by the function
\[ F(r)=a_0 + a_1 S_{m,R_e}(r), \]
where $S_{m,R_e}(r)$ represents the set of convolved model profiles,
$a_0$ is the background level, and $a_1$ yields the cluster brightness.
Generally the best fitting $R_e$ agreed to within 0.2 pixels for the two
sets of models.  Here we report the mean best fitting $R_e$.  Typical
external errors are $\sigma_{R_e} \approx 0.5$ pixels, and
$\sigma_{m_{220}} \approx 0.2$ mag as determined from comparing the
measurements of the two separate images of NGC~1705.  Measurements of 65
stellar profiles in archived frames of the Galactic clusters NGC~104 and
NGC~188 (taken with the same setup) were used to determine the
resolution limit of our technique.  We find that sources with $R_e \leq
1.52$ pixels can not be distinguished from stars at a better than 90\%\
confidence.  The objects in NGC~5253 and NGC~1705 that were selected as
likely point sources are all unresolved by our technique.

The resultant $R_e$ are shown as a function of $M_{220}$ in
Fig.~{14}.  The measurements are tabulated in
Table~\ref{t:sources}, which also lists the position offsets (relative
to the coordinates given in Table~\ref{t:obslog}) and magnitudes of
the sources.  Sources are included in this list if they had their
profile fitted, or if they are bright enough, or if they are discussed
earlier in the text. The brightness limit is $\Muv < -14$ for most
host galaxies, and $\Muv < -16$ for NGC~3690 and NGC~7552 because of
their large UV absorptions, $A_{220}$, due to dust extinction
(Table~\ref{t:redex}).  Note that $A_{220}$, (derived in
\S\ref{ss:red}), is applied uniformly to all sources in a galaxy (i.e.\
assuming a uniform foreground screen).  If the dust distribution is
non-uniform and the sources evenly distributed, then they will
preferentially be detected where the dust is thin, and \Muv\ may be
preferentially underestimated.  The apparent magnitude \muv\ of the
compact sources is from the aperture photometry, while \Muv\ is that
derived from the profile fitting (where applicable).

The resolved objects have $R_e$ ranging from 0.6
to 22 pc.  This is nearly identical to the range of $R_e$ found in
Galactic globular clusters (van den Bergh et al.\ 1991).  Thus the
sizes of the compact sources are consistent with globular cluster
sizes.  However the distribution of sizes for the compact sources
differs from globular clusters, with clearly a larger fraction of our
sources at the high $R_e$ end; 14\%\ of our sample have $R_e > 10$ pc,
while only 3\%\ of the Galactic globular clusters are this large.  In
addition the resolved sources appear to follow a $R_e \propto L^{1/2}$
correlation (i.e.\ constant surface brightness), which is not seen in
the Galactic globular cluster sample.

There are a number of reasons why we should be wary of these results.
(1) 42\%\ of the sample have only upper limits in $R_e$. (2) There is
no correlation of $R_e$ with $L$ within a galaxy (in NGC~3991 the
quantities are anti-correlated).  (3) There is a strong Malmquist
bias in our sample: for galaxies at large distances ($D$) we can not
see the faintest clusters, and at small $D$ there are very few of the
highly luminous SSCs.  The exception is NGC1705-1 which is saturated
in our images.  (4) Many of the sources at large $D$ appear elongated,
and thus may have faint neighbors.  (5) The size of the fitted region
increases with $D$.  Since clusters are preferentially found embedded
in the highest surface brightness background the chance of
contamination increases as $D^2$ especially if the $\mu$ profile is
peaked towards the cluster (e.g.\ NGC~1705, NGC3690-Aa,Ab; see
fig~{5}).

To help compensate for the Malmquist bias, we supplemented our sample
with measurements of NGC1705-1 and the two SSCs in NGC~1569 ($D = 2.2$
Mpc) using the planetary camera (PC) images presented by (O'Connell
et al.\ 1994).  Profiles from their summed F555W and F785LP images were
measured with a technique nearly identical to that used on the FOC
data.  The only difference was that the smallest annulus was not
discarded.  We find $R_e =$ $1.1\pm 0.3$, $1.7 \pm 0.2$, and $1.2 \pm
0.2$ pc for NGC1705-1, NGC1569-1, and NGC1569-2 respectively.  These
are the mean $R_e$ in the two bands, with the error representing half
the difference between the bands.  The sizes for these nearby SSCs are
shown in Fig.~{14}.

To test the effects of contamination we simulated images of NGC~1705
at $D$ ranging from 6.2 to 49 Mpc.  The intrinsic image was taken to
be the Lucy deconvolved frame of NGC~1705b, which had NGC1705-1
removed prior to deconvolution.  NGC1705-1, was then added back to
this image as a circularly symmetric Gaussian profile with $R_e = 1.1$
pc as measured in the PC images.  The intrinsic image was rebinned to
simulate the intrinsic image at different $D$, and then reconvolved
with the PSF.  The cluster profiles from the resulting images were
then measured.  We find that at its observed luminosity NGC1705-1,
would remain unresolved out to 49 Mpc.  However, we would measure $R_e
= 6.9$ pc for a cluster half as bright on top of the same background
at a distance of 37 Mpc.  Thus, the amount of contamination from the
background depends critically on the contrast of the cluster against
the background.

We conclude that the size estimates of the more distant clusters
should be treated with caution. Although some of these objects may
indeed be isolated clusters with $R_e \approx 10$ pc, the data are not
sampled well enough to measure smaller sizes, and $R_e$ could easily
be contaminated with high surface brightness structure correlated with
the clusters.  To be safe one should not try to measure $\Wfo = 2R_e$
smaller than the Nyquist width.  So to unambiguously measure clusters
with $R_e = 1$ pc with the HST one should not look at galaxies beyond
$D = 9$ Mpc.

NGC~1569, NGC~5253, and NGC~1705 meet this criterion.  The resolved
sources in these galaxies have $R_e \leq 2.9$ pc, and show no
correlation of $R_e$ with $L$.  The median of the clusters in this
sub-sample is 1.3 pc, smaller than the Galactic globular cluster
sample of van den Bergh et al.\ (1991) which has a median $R_e = 2.6$
pc. This may not represent a real difference since the model surface
brightness profiles of van den Bergh et al.\ (1991) have more power at
large radii than a Gaussian.  Recently O'Connell et al.\ (1995) have
measured the sizes of SSCs in M82 ($D = 3.6$ Mpc) and find a mean
$\Wfo \approx 3.5$ pc from deconvolved HST images.  This translates to
$R_e \approx 1.8$ pc if a Gaussian profile is assumed for the
clusters.  Therefore the accumulated evidence on nearby starbursts is
that the $R_e$ of SSCs is similar to that of typical globular
clusters.

Whitmore et al.\ (1993) and WS measured cluster $R_e$ in the merging
systems NGC~7252 and NGC~4038/4039 using WFPC-1 data.  They also adopt
a Gaussian model for the cluster profiles, but derive $R_e$ from only
one data point for each object: the magnitude difference between the
light measured in apertures of $r = 0.5$ and 3 pixels.  They measure
typical $R_e \approx 7-20$ pc for the clusters in these systems,
assuming revised distances of $D = 65, 19$ Mpc respectively ($H_0 =
75\, \kms\,{\rm Mpc^{-1}}$). These sizes are at the large end of the
size distribution of Galactic globular clusters as pointed out by van
den Bergh (1995).  However NGC~7252 and NGC~4038/4039 are too distant
to reliably measure $R_e$ as small as a few pc and the caveats
concerning our measurements at large $D$ also apply.  In particular
the area they fit (out to $r = 27-40$ pc) may be contaminated by
surrounding high surface brightness structure.

\subsection{Luminosity Function.}

Figure~{15} shows the combined \Muv\ luminosity function (LF) of
SSCs excluding those in NGC~7552 and NGC~3690 (because of their large
extinction corrections and corresponding bright limiting magnitude).
The \Muv\ magnitudes produced by the profile fitting are somewhat
brighter than than those from the aperture photometry because most
SSCs are resolved.  To avoid biasing the LF, the sources with just
aperture photometry have \Muv\ in Table~\ref{t:sources} corrected by
the median in $\Muv({\rm fit}) - \Muv({\rm ap.\ phot}) = -0.26$ (range
is from --1.7 to 0.4 mag) before constructing the LF.

The luminosity function is rising out to the faint magnitude limit of
$\sim -14$ with no sign of decrease.  Because of the low total number
of sources in the LF, and the differing distances and limiting
magnitudes of the sample we do not fit the LF.  Instead we compare it
to a power law LF: $\phi(L)dL \propto L^\alpha dL$, where $\phi(L)dL$
are the number of clusters with luminosities between $L$ and $L + dL$,
and $\alpha = -2$.  This slope is chosen because it is representative
of the $\alpha$ of several other systems of young clusters.  For
example Elson \&\ Fall (1985) find $\alpha = -1.5$ for LMC clusters.
The LF they present for Galactic open clusters was not fit but appears
to be a power law with a steeper slope. WS find the SSC LF in the
merging system NGC~4038/4039 are also well fit by a power law LF with
$\alpha = -1.78 \pm 0.05$. Kennicutt et al. (1989) find that the
\Halpha\ LFs of \HII\ regions in spiral galaxies are well fit by a
power law with $\alpha = -2 \pm 0.5$.  This is also presumably the LF
slope of the ionizing clusters within each \HII\ region.  Thus systems
of young clusters, including SSCs, appear to be well represented by a
power law LF with $\alpha \approx -2$.

\subsection{The nature of SSCs}

There is much speculation in the literature as to whether SSCs
represent proto-globular clusters (e.g.\ MFDC; Whitmore et al.\ 1993;
WS), or are merely high luminosity open clusters (e.g.\ van den Bergh,
1995).  We do not wish to enter a semantics debate on the distinctions
between open and globular clusters (c.f.\ Stetson, 1993).  A better
way to express the question is would SSCs resemble globular clusters
if left to evolve?  This is especially relevant to elliptical
galaxies.  The hypothesis that these are formed by mergers of disk
galaxies (Toomre, 1977) which result in the creation of SSCs may
explain the high specific frequency of globular clusters around E
galaxies (Ashman \&\ Zepf 1992).  To answer this question it is
important to determine whether the ensemble properties of SSCs are
consistent with globular clusters.

Our results indicate that the range in luminosity is consistent with
that expected for young globular clusters if they had an IMF extending
up to $m_u \approx 100$ \Msun.  Similarly the range in $R_e$ is
consistent with the present range seen in globular clusters.  However
the difficulties come when trying to compare the detailed distribution
of $R_e$ and $L$.  As discussed above, when only the nearest SSCs are
considered, the $R_e$ of globular clusters and SSCs are consistent.
The projected effective mass density of NGC1705-1 (the densest cluster
for which we have a reliable size) is $\Sigma \gapeq 5 \times 10^4\,
\Msun\, {\rm pc^{-2}}$ if its stellar IMF extends down to $m_l = 0.1$
\Msun. This limit (corresponding to the peak $L_{220}$ of
an SSC) is similar to the projected central density of M80 and 47 Tuc
$\Sigma \approx 7 \times 10^4\, \Msun\, {\rm pc^{-2}}$ (Peterson \&\
King, 1975).

The available LF of SSCs are well characterized by power laws with
slope $\alpha \approx -2$, and are not consistent with that of present
day globular clusters.  The latter have an approximately Gaussian
luminosity function in the $N$ versus $M_V$ plane (Harris, 1991).  The
globular cluster luminosity function is essentially a mass function,
since the age spread amongst globular clusters is small compared to
their mean age.  This is not the case for SSCs which are still being
formed in their host starbursts.  Meurer (1995b) shows how the power
law SSC LF in the Antennae system is well modeled by a globular
cluster {\em mass\/} function and continuous cluster formation.  This,
of course, does not mean that a Gaussian mass function is the only
mass function that can reproduce the power law LF of SSCs.  Even if
the SSC mass function starts as a pure untruncated power law, it is
conceivable that a lower mass truncation could be imposed by
destruction mechanisms (Fall \&\ Rees, 1977; see however, Fall \&\
Rees, 1988). Then the SSC LF may evolve into a Gaussian if the SSCs
are placed in the right environment.  All in all, the LF alone does
not provide a convincing constraint on the nature of SSCs.

At this stage, the known properties (mainly size and luminosity) of
SSCs are consistent with the hypothesis that they represent
proto-globular clusters.  There are other properties that can be
measured to further test this hypothesis.  Spectroscopy, colors and
extinction measurements would be useful in determining the ages for
SSCs.  From these it might be possible to construct a true mass
function for the SSCs.  One problem with the proto-globular cluster
hypothesis is that we don't even know if the stellar mass functions of
globular clusters and SSCs overlap; our SSC observations are only
sensitive to stars with $m_* > 5$~\Msun, while the only stars
presently in globular clusters have $m_* < 1$~\Msun.  Since most of
the mass in the IMF should come from the low mass stars, mass
estimates of SSCs will be crucial in determining if they could evolve
into globular clusters.  The measurements will be difficult since
globular clusters typically have velocity dispersions $\sigma \lapeq
10$ \kms\ (Peterson \&\ King, 1975).  Finally, it would also be good
to have measurements of the metallicities of SSCs to see if they agree
with globular clusters.

\section{Summary: the anatomy of starbursts}\label{s:summ}

High resolution ultraviolet images, combined with HST and IUE
spectroscopy have allowed us to dissect the structure of starburst
galaxies covering a large range of luminosities and morphologies.
Despite their differences we can now begin to describe the anatomy of
a starburst.

{\em Dust distribution.\/} A significant component of the dust content
of starburst galaxies must be in a foreground screen near the starburst.
This is deduced from the observed strong correlation of the infrared to
UV flux ratio $\log({\em FIR}/F_{220})$ with the UV spectral slope,
$\beta$, for our sample and other galaxies detected with both the IUE
and IRAS satellites.  This correlation shows that as the spectral energy
is redistributed from the UV to the far infrared, the spectrum is
reddened.  The range in $\beta$ over which the correlation applies rules
out geometries were the dust is mixed evenly with the UV emitting stars
or in a very clumped distribution.  Only the foreground screen geometry
adequately models this range.  The average ratio of 60$\micron$ and
$100\micron$ flux suggests that the screen is located $\sim 2.5 R_e$
from the starburst center.

It is also clear that not all the dust is in a uniform foreground
screen.  Our UV images clearly show dust lanes in some cases
(NGC~5253, NGC~3310, NGC~7552).  Data from other wave bands indicate
that the extinction distributions in the most highly reddened
galaxies, NGC~3690 and NGC~7552 are not uniform.  Nevertheless, even
in these cases the observed reddened values of $\beta$ indicate that
foreground dust has a large covering factor and may be responsible for
a large fraction of the far-IR flux.

{\em Overall morphology.\/} The UV morphology of a starburst is highly
irregular, composed of one or several diffuse, usually oddly shaped,
cloud(s) with embedded compact sources.  The size and luminosity of
the brightest compact sources indicate that they are star clusters.
The three nearest galaxies in our sample (NGC~5253, NGC~1705, IZw18)
are near enough to resolve into stars which are spread throughout the
diffuse component, suggesting that most of the diffuse light is
unresolved stars and not scattered light.  The radial profiles are not
well parameterized by either an exponential or an $r^{0.25}$ law.

The size of the clouds allow the star formation duration to be
constrained: it is unlikely that simultaneous star formation is
occurring over extended regions on a timescale shorter than the
crossing time.  This is typically $\sim 10$~Myr for typical gas
velocities observed in starbursts.  The timing arguments are less
stringent for the clusters each of which are expected to form over
timescales $\Delta t < 0.1$ Myr (Larson, 1988).  The star formation
history is then likely to be more or less continuous over timescales
of $\gapeq 10$ Myr for the diffuse component, with the clusters
representing true staccato ``instantaneous'' starbursts.

Most starbursts (7/9 galaxies) are observed with a preferred effective
surface brightness, $\langle\mu_{e,0}\rangle$ $= 15.9 \pm 0.4$.  The
corresponding effective star formation rate per area in high mass stars
(5 -- 100 \Msun) is $\msfae \approx 0.7$ \Msun Kpc$^{-2}$ yr$^{-1}$ for
a Salpeter (1955) IMF slope (or ${\rm SFA_e} = 3.5\, \Msun$ ${\rm yr^{-1}\,
Kpc^{-2}}$ if the IMF extends down to 0.1 \Msun).  Since we have
selected by UV brightness we may be preferentially selecting galaxies
with a high \msfae, at least compared to ``normal'' galaxies.  Thus
$\langle\mu_{e,0}\rangle$ may represent the upper limit to the UV
intensity that one can find in a galaxy. Similar results are found in
the far IR by Armus et al.\ (1990) and Lehnert (1992). These results
suggest that there is a negative feedback mechanism limiting the
\msfae\ and thus the UV radiation field.  What this mechanism is, is
not apparent from our observations, but there are numerous
possibilities including mechanical heating from SNe and winds, cosmic
ray heating and the intense UV radiation field itself.

{\em Cluster Formation.\/} UV bright star clusters are found within
all starbursts, and very few are found outside of them.  Young
luminous clusters are increasingly being found with the HST; examples
are noted in \S\ref{s:intro}.  The host galaxies all have starburst
characteristics.  Thus cluster formation is an important mode of star
formation in starbursts.  Clusters are manufactured at a high
efficiency.  On average $\sim 20$\%\ of the UV light comes from
clusters.

The fraction of light in clusters is (weakly) correlated with
$\mu_{e,0}$; galaxies with the most intense UV field have more light
in clusters.  As one increases the UV field strength, more of the star
formation is in clusters.  This is also seen locally; 90\%\ of the
brightest clusters, the ``super star clusters'' or SSCs, are found to
have an underlying $\mu_{220}$ within 1.5 mag arcsec$^{-2}$ of the
most intense found in the starburst.

Thus SSCs are formed in the heart of a starburst.  This suggests that
cluster formation is related to the mechanism that regulates $\mu_{e,0}$
of the whole starburst. The observations summarized by Larson (1993)
indicate that cluster formation is triggered by neighboring star
formation.  Perhaps clusters may inhibit further star formation, at
least in the immediate surroundings.  Indeed the effect of cluster
formation on the ISM can be much more dramatic than a 20\%\ contribution
to $L_{220}$.  For example Malumuth \&\ Heap (1994) shows that a simple
census of the ionizing stars in the cluster NGC~2070 in the 30 Doradus
nebula alone can account for 40\%\ of the ionizing flux of the whole
nebula which is $\sim 1$ Kpc across.  Another example is NGC1705-1,
which is probably the dominant energy source of NGC~1705's spectacular
galactic wind which extends out to the Holmberg radius (Meurer et al.\
1992; Meurer 1989).

{\it The nature of SSCs.\/} The ranges in both half light radius $R_e$
and UV luminosity $L_{220}$ of SSCs are consistent with those expected
for young ``proto-''globular clusters.  The median $R_e$ of SSCs with $D
< 9$ Mpc is consistent with the $R_e$ of present day globular
clusters. We have shown that the $R_e$ measurements of more distant SSCs
(e.g.\ ours and those of Whitmore et al.\ 1993; Whitmore \&\ Schweizer
1995) are likely to be contaminated by the high surface brightness
structure correlated with the presence of SSCs.  The SSC luminosity
function (LF) is well characterized by a power law with slope $\alpha
\approx -2$ (this work and Whitmore \&\ Schweizer, 1995) which is
similar to the LFs of Galactic open clusters and LMC clusters (Elson \&\
Fall 1985) and \HII\ regions (Kennicutt et al., 1989) but not that of
Galactic globular clusters (Harris, 1991).  Because SSCs are not coeval
whereas Galactic globular clusters are (in effect), the SSC LF may be
consistent with a globular cluster mass function (Meurer, 1995b).

Thus SSCs remain plausible proto-globular cluster candidates.  Mass
and metallicity estimates of SSCs could confirm whether SSCs are
indeed proto-globular clusters.  If so, then the globular cluster
forming clouds in the Searle \&\ Zinn (1978) model of galaxy formation
may have looked like starbursts.  It is interesting that on purely
theoretical grounds Larson (1988) came to the same conclusion.

\acknowledgements

{\em Acknowledgements:} We thank Antonella Nota, Perry Greenfield, Dave
Baxter, Warren Hack, Guido De Marchi, and Bernie Simon for discussion on
the intricacies of the FOC and its data products.  Lee Armus, Duncan
Forbes, Matt Lehnert and Abi Saha kindly lent us optical CCD images for
comparison.  We thank Gustavo Bruzual for providing us with his GISSEL
models.  We thank Daniela Calzetti, Michael Fall, Nino Panagia, and Brad
Whitmore insightful and stimulating discussions which helped shape this
paper.  We are grateful for the suggestions of the anonymous referee.
We are grateful for the support we received from NASA through grant
number GO-3591.01-91A, from the Space Telescope Science Institute, which
is operated by the Association of Universities for Research in
Astronomy, Inc., under NASA contract NAS5-26555, and NASA grants
GO-4370.03-92-A, and NAGW-3138, administered by The Johns Hopkins
University. DRG acknowledges support from Hubble Fellowship award
HF-1030.01-92A, provided by NASA and STScI.  Literature searches were
performed using NED, the NASA\-IPAC Extragalactic Database, a facility
operated by the Jet Propulsion Laboratory, Caltech, under contract with
the National Aeronautics and Space Administration.

\appendix

\section{Ultraviolet properties of young stellar populations}

\subsection{Models}

In order to translate raw measurements of UV properties of starbursts
into physical quantities such as mass, reddening, and star formation
rate, we have measured modeled spectra of stellar populations.  We
primarily employ spectra generated from the code of Leitherer and
Heckman (1995; hereafter LH) with ages from 1 to 100 Myr.  A power law
IMF with a Salpeter (1955) slope of $\alpha = 2.35$ (eq.\ 2 of LH)
with mass limits $m_l = 1.0$ \Msun\ to $m_u = 100$ \Msun\ was adopted.
We refer to these parameters as the ``standard'' IMF.  Two star
formation histories were considered: an instantaneous star burst (ISB)
and a constant star formation rate (CSF).  All models employ solar
metallicity evolutionary tracks and stellar atmospheres.  The broad
band properties examined here are not significantly affected by
metallicity.  The LH models calculate both the stellar and nebular
continuum contributions to the total flux.  We consider the stellar
and total (= stellar + nebular continuum) spectrum separately.

We supplemented the LH models with models from the Bruzual \&\ Charlot
(1993) Galaxy Isochrone Synthesis Spectral Evolutionary Library
(GISSEL).  The GISSEL models also have a solar metallicity and a
Salpeter IMF, but with mass limits $m_l = 0.1$ \Msun\ to $m_u = 125$
\Msun.

The ISB models have been normalized to an initial mass of
$10^6$~\Msun\ formed over the standard IMF.  The CSF models are
normalized to a star formation rate of $1 \Msun\, {\rm yr^{-1}}$ over
the same mass range. Note that in both cases the GISSEL models still
include the flux of the stars outside the LH mass range, although the
extra UV flux is negligible.

\subsection{Measurements}

Broad band magnitudes, $m_\lambda$, were extracted with the SYNPHOT
package in IRAF  using the STMAG form:
\[
m_\lambda = -2.5 log \left( \frac{\int f_\lambda(\lambda) T(\lambda)
\lambda d\lambda}{\int T(\lambda) \lambda d\lambda} \right) - 21.1,
\]
where $f_\lambda(\lambda)$ is the flux density in erg cm$^{-2}$
s$^{-1}$ \AA$^{-1}$, and
$T(\lambda)$ is the dimensionless passband throughput, in this case
defined by the combined throughput of the HST optics (pre-COSTAR), the
FOC F/96 camera, and the F220W filter.  This
$T(\lambda)$ has a pivot wavelength $\lambda_c =2320$\AA\ and FWHM $=\Wfo =
580$\AA.  The mean flux density is then
\[ \langle f_{220} \rangle = 10.0^{-0.4(\muv + 21.1)}\, \FUA.\]
We define $F_{220}$ and $L_{220}$ to be
\[ F_{220} = \lambda_c \langle f_{220} \rangle \]
\[ L_{220} = 4 \pi D^2 F_{220}. \]
A source with a 220 luminosity equal to the sun's Bolometric
luminosity, that is with $L_{220} = L_{\rm \sun,Bol} = 3.83 \times 10^{33}
\, {\rm erg\, s^{-1}}$ (Allen, 1973) will have an absolute magnitude
in this system of $\Muv = 3.55$.

Figure~{16} shows the temporal evolution of various quantities
of interest.  Panels on the left are for ISB populations, those on the
right are for CSF populations. It is immediately apparent that the LH
and GISSEL are in good agreement.  The rapid fluctuations in the LH
ISB models in some panels are an artifact of insufficient mass
resolution in the models generated here.  \Muv\ is the absolute
magnitude. $\beta$ is the UV spectral slope and is extracted by
fitting the spectra with $\log(f_\lambda) = C +
\beta\log(\lambda)$ using the points in the ten spectral windows
defined by Calzetti et al.\ (1994; hereafter C94).  The relatively low
spectral resolution LH and GISSEL models were first interpolated to a
2\AA\ pixel size, to make sure there were sufficient data points in
each window. The color $220-V$ ($= \muv - m_V$) is presented as a
tie in to other colors which are calculated by LH and Bruzual and
Charlot (1993).  The redleak parameter, $RL$, is defined as $RL =
\muv({\rm square}) - \muv$, where $\muv({\rm square})$ is $m_\lambda$
for a square $T(\lambda)$ having a central wavelength $\lambda_c =
2320$~\AA, and width $\Delta\lambda = 580$~\AA.  The parameter, $NC =
\muv({\rm stars}) - \muv({\rm total})$, represents the amount of
contamination of the total light by nebular continuum.  The evolution
of $\log(L_{220}/L_{\rm T})$, where $L_{\rm T}$ is the total
luminosity, was also calculated but not shown in this figure (the
evolution is similar to that of $220- V$).

Figure~{16} also shows the mass to light ratio, $M/L_{220}$.  The
numerator represents the total {\em initial\/} mass of the stars.
This is not the same as the instantaneous mass in stars since a large
fraction of the initial mass is returned to the ISM via stellar winds
and SNe.  LH provide detailed calculations of the rate of mass return
to the ISM.

The $M/L_{220}$ results are for the standard IMF which extends down to
1 \Msun.  So what mass stars are F220W measurements most sensitive to?
This was estimated by using the LH code to construct CSF models
at $t=10$ Myr (a typical ionizing population), in which $m_u$ varies
between 3 and 120 \Msun, while the number of stars with $m_* < 3
\Msun$ is held constant.  Figure~{17} shows the change in \muv\
(stellar component only) as a function of $m_u$, relative to the
standard IMF.  Thus for the standard IMF, 50\%\ of the F220W light is
produced by stars with $m_* < 28$\Msun, and only 5\%\ by the stars with
$m_* < 8~\Msun$.  If $m_u = 30~\Msun$ then the 50\%\ and 5\%\ levels
are reached for $m_* < 18, 6$~\Msun\ respectively.  So for the
standard IMF slope and $30 \leq m_u/\Msun \leq 100$, the typical emitting
star has $m_* \approx 20\Msun$, and stars with $m_* < 5$~\Msun\
produce negligible emission.

With the standard IMF, the luminosity is mostly produced by the high
mass stars while the mass is heavily weighted towards the low mass
stars.  Therefore, the mass to light ratio is highly dependent on
$m_l$.  If $m_l = 5$~\Msun\ or 0.1~\Msun, then the $M/L_{220}$ in
Fig.~{16} should be multiplied by 0.46 or 2.55 respectively.

Figure~{16} shows that NC drops to zero by $t = 10$ Myr for
the ISB models.  This indicates that by this age there is no ionizing
flux for ISB models.  Since starbursts display a recombination
spectrum, they must have $t < 10$ Myr if they were formed in a true
instantaneous burst.  If the CSF model is appropriate they can have
any age, although in this paper we will only consider $t < 100$ Myr.
Figure~{16} shows that $\beta < -2.2$ and {\em RL\/} is
insignificant for ionizing populations.  It is also clear that RL is
insignificant for unreddened ionizing stellar populations. In
Table~\ref{t:avgq} we present time averages, and ranges for $220-V$,
$RL$, $NC$, $\beta$, $\log(L_{220}/L_{\rm T})$, and $M/L_{220}$ (with three
different $m_l$), for ionizing populations.  These results are for the
total flux (stellar plus nebular continuum) models of LH.  The
averages are for equally spaced points in $log(t)$ over the range $6
\leq log(t/(1 {\rm yr})) \leq 7$ for the ISB models and $6 \leq
log(t/(1 {\rm yr})) \leq 8$ for the CSF models.

\subsection{Reddening and extinction}

We examined how \muv\ and $\beta$ are affected by four extinction
laws: (1) the Galactic law of Seaton et al.\ (1979); (2) the LMC law of
Howarth (1983); (3) the starburst law of Kinney et al.\ (1994; K94);
and (4) the starburst law of C94\footnote{The form of the K94 and C94
extinction laws is that applicable to the continuumn.}.  We explore
their effects on CSF models with ages 1, 10, and 100 Myr. Tests show
that our results also hold for ISB models with $t < 10$ Myr.

We define UV ``color excess'' to be
\[ E(\beta) = \beta - \beta_0, \]
where $\beta_0$ is the unreddened UV slope.  The relationship between
$E(\beta)$ and \ebv\ is extremely well fit by a linear relationship:
\[ E(\beta) = r_\beta \ebv. \]
The values of $r_\beta$ for the four reddening laws are given in the
second column of Table~\ref{t:redprop}. The residuals of the fits
($E(\beta) - r_\beta \ebv$) have an rms $\leq 0.0001$ in all cases.
That the UV and optical color excesses should be related by a single
slope makes sense, since the strongest terms in the reddening laws are
essentially power laws.  However, the Galactic and LMC extinction law
have a significant bump at $\lambda = 2175$\AA\ which is thought to be
due to small grains perhaps with graphite cores (Mathis, 1994).  The
continuum regions used to define $\beta$ avoids the peak of this bump,
but not its wings.  The values of $r_\beta$ are decreased due to the
wings of the bump for the LMC and (especially) Galactic laws.
Therefore, for these laws, the value of $r_\beta$ depends critically
on which continuum regions are adopted. The values quoted here are
only for the continuum regions of C94.

We parameterize the effects of extinction on the UV fluxes in terms of
the ratio of total to selective extinction $X_\lambda$:
\[ X_{220}     = \frac{A_{220}}{E(B-V)} \]
where $A_{220} = m_{220} - m_{220,0}$ is the extinction and
$m_{220,0}$ is the intrinsic magnitude. This ratio is shown as a
function of \ebv\ for the three representative CSF models, and the
four extinction laws in Fig.~{18}.  $X_{220}$ declines with
\ebv\ because of the broad passband; as \ebv\ increases the effective
wavelength shifts to the red where $X_\lambda$ is lower.

For each reddening law, simple polynomial fits to $X_{220}(E(B-V))$
were made:
\[ X_{220} = \sum_{i=1}^n a_i E(B-V)^i, \]
The highest order used was $n=2$ or 3.  The polynomial coefficients
$a_i$  and the rms of the fits are reported in
Table~\ref{t:redprop}.

The four extinction laws are very similar in the optical but diverge
markedly in the UV.  It is better to parameterize the UV extinction
curves in terms of the local (UV) color excess.  The bottom panel
of Fig.~{18} shows
\[ X'_{220} = \frac{A_{220}}{E(\beta)} = X_{220}/r_\beta, \]
as a function of $E(\beta)$.  To simplify the diagram, we show only
the fits to $X_{220}$ after transforming coordinates from \ebv\ to
$E(\beta)$.  The $X'_{220}$ curves have a much simpler behavior than
the $X_{220}$ curves.


\begin{flushleft}

\end{flushleft}

\clearpage

\pino

{Fig.~{1}} -- Grey scale representation of the reduced
FOC images. Panels are as follows: (a) NGC~1705 frame a; (b) IZw18; (c)
NGC~3310; (d) NGC~3690; (e) NGC~3991; (f) NGC~4670, (g) NGC~5253; (h)
Tol1924-416; (i) NGC~7552.  The images are reproduced with a square
root stretch with the exception of the NGC~7552 image which uses a
cube root stretch. In all panels the long and short arrows point north
and east, respectively, and have lengths of 2\as\ and 1\as.

{Fig.~{2}} -- FOC $m_{220}$ measurements compared to
those from IUE.  The dotted line marks where they are equal.

{Fig.~{3}} -- Central region of NGC~1705.  Top:
NGC~1705a frame before subtracting model of NGC1705-1; bottom: after
subtraction of NGC\-1705-1.  The arrows indicate the orientation and
scale.  The long arrow is 1\as\ long and points to the north, the
shorter 0.5\as\ long arrow points to the east.

{Fig.~{4}} -- Schematic representation of the contents
of each frame, after rotating so north is up and east is to the left.
The $\Delta\alpha$, $\Delta\delta$ coordinates are the offsets to the
east and north respectively from the coordinates listed in Table 3.
The FOC frame is shown as the large tilted square, while the shaded
areas are the regions masked out of the image when determining the
surface brightness profile.  The contours indicate isophotes in the
smooth component images.  They are drawn from the sequence 0.25\%,
0.5\%, 1.0\%, 2.5\%, 5\%, 10\%, 25\%, 50\%, 75\%, 90\% of the peak
surface brightness in the smooth component image.  The compact sources
are marked by overplotted solid circles and crosses.  The size of the
circle decreases linearly with \muv, while the cross size remains
fixed.  The correspondence between \muv\ and symbol is given in panel
a.  The galaxies, the faintest isophote shown (in mag arcsec$^{-2}$),
and the corresponding percentage of the peak are: (a) NGC~1705a,
20.08, 1\%; (b) NGC~1705b, 20.05, 1\%; (c) IZw18, 21.95, 1\%; (d)
NGC~3310, 18.52, 10\%; (e) NGC~3690, 20.80, 2.5\%; (f) NGC~3991,
21.27, 1\%; (g) NGC~4670, 20.64, 1\%, (h) NGC~5253, 19.33, 5\%; (i)
Tol1924-416, 21.38, 1\%; and (j) NGC~7552, 21.42, 0.25\%.  **** NOTE
CONTOURS NOT SHOWN IN FTP FILE fig04.ps! ****

{Fig.~{5}} -- Surface brightness profiles extracted
from the images.  The thick lines are profiles derived from the total
image, while the thin lines connecting the dots shows the profiles
derived from the smooth component images.  The dotted lines show the
effect of changing the ``sky'' level by $\pm 1\sigma_{\rm sky}$ on the
smooth component profile.

{Fig.~{6}} -- The far infrared excess ${\rm IRX} \equiv$\linebreak
$\log({\it FIR}/F_{220})$ plotted against UV spectral slope,
$\beta$. The $F_{220}$, and $\beta$ have been corrected for Galactic
foreground extinction using the Seaton (1979) extinction law (see
appendix).  Galaxies in our sample are indicated by solid circles.
Fluxes are tabulated in Table 8. The downward pointing arrow marks the
upper limit of IRX for IZw18 which was undetected by IRAS.  Additional
galaxies from K93 detected by both IRAS and IUE are indicated by
crosses. The range $\beta_0 = \pm -2.5$ for unreddened ionizing
populations (see appendix) is shown as the hatched region. The curves
show the relationship for the near foreground screen models discussed
in the text and four extinction laws: Galactic (solid line), LMC
(dotted), and the starburst extinction laws of C94 (dashed), and K94
(dot dash).

{Fig.~{7}} -- $\beta$ plotted against Balmer decrement
determined reddening, $E(B-V)_{\rm BD}$.  The effects of reddening a
$\beta_0 = -2.5$ spectrum using the extinction laws of K94 (assuming
that $E(B-V)_{\rm UV} = E(B-V)_{\rm BD}$) is shown with the dot-dashed
line (appendix).

{Fig.~{8}} -- Global properties of the sample.  Panel (a)
plots the effective surface brightness in terms of the effective
massive star formation rate per unit area, \msfae\ (see text for
definition), against UV luminosity, $L_{220}$. Panel (b) plots \msfae\
against the effective radius of the starburst, $R_e$. Panel (c) plots
$L_{220}$ against $R_e$. Panel (d) plots the extinction corrected UV
flux $F_{220,0}$ against the corresponding \Halpha\ flux, $F_{\rm
H\alpha,0}$.  The units are \msfae: $\Msun\, {\rm Kpc^{-2} yr^{-1}}$,
$L_{220}$: ${\rm erg\, s^{-1}}$, $R_e$: pc, and $F_{\rm H\alpha,0}$,
$F_{220,0}$: \FU.  Data points are represented with filled circles
except for: (1) NGC~3690, shown as squares, all three components are
shown in panels (a--c), while the integrated fluxes are shown in panel
(d); and (2) NGC~1705, the solid pentagon includes the light of
NGC1705-1, the open pentagon excludes it. The two symbols for the
NGC~1705 are connected by a solid line.  The dotted line in panel (c)
shows a line of constant surface brightness.  The dotted, dashed, and
dot-dashed lines in panel (d) are the expected correlations for
constant star formation rate populations of ages 1, 10, and 100 Myr
respectively from the models of LH, assuming a Salpeter IMF over the
mass range 1 -- 100 \Msun.  Note that in panel (d) the UV fluxes are
dereddened by \ebvuv, while the \Halpha\ fluxes are dereddened by
$E(B-V)_{\rm BD}$.  Tables 2, 8, and 9 contain the quantities used to
construct this figure.

{Fig.~{9}} -- Histograms of compact source absolute
magnitudes, \Muv.  The bottom scale on each panel shows the absolute
magnitude where the full extinction correction, $A_{\rm 220,T}$, has
been applied, while the top scale shows the scale where only the
Galactic extinction $A_{\rm 220,gal}$ has been removed ($A_{220}$
values are given in Table 2). Note, this is the only figure that shows
\Muv\ with just the $A_{\rm 220,gal}$ correction.  All remaining
figures show \Muv\ of the sources with the full extinction correction.
For the NGC~1705 histogram, objects detected in both frames are given
a weight of 1, and those detected in only one frame have a weight of
0.5.

{Fig.~{10}} -- Fraction of UV light in the clumpy image,
$f_{\rm clumpy}$ plotted against \msfae.  Symbols are the same as in
in fig.~{8}.

{Fig.~{11}} -- Absolute magnitude, \Muv, versus local
underlying surface brightness, $\mu_{220,0}$, for all compact
objects. $\mu_{220,0}$ was extracted from the average surface
brightness in the smooth component image within a $0.34'' \times
0.34''$ box centered on each object. \Muv\ is from radial profiles
fits for those objects marked with an open square, and are from the
aperture photometry for all other sources.  Objects identified in only
one of the NGC~1705 frames are marked with a cross. The top axis
converts $\mu_{220,0}$ to the corresponding massive star formation
rate per unit area, MSFA.

{Fig.~{12}} -- Panel a shows \Muv\ versus $\mu_{220,0}$
for the SSCs in the lightly reddened starbursts NGC~1705 (closed
pentagon), NGC~3310 (closed circles), NGC~3991 (open stars), NGC~4670
(closed triangles), and Tol1924-416 (open squares). The upper scale
on panel a converts $\mu_{220,0}$ to the massive star formation rate
per unit area, MSFA.  Panel b shows the histogram of number of these
clusters as a function of $\mu_{220,0}$.  Panels c and d are the same
as a and b except $\mu_{220,0} - \mu_{\rm 220,min}$ is plotted as the
abscissa.

{Fig.~{13}} -- Examples of compact source radial profile
fits. Those on the left side are resolved, those on the right are
essentially unresolved ($R_e \leq 1.52$ pixels).  $S(r)$ is the mean
count level in circular annuli. The errorbars are equal to $\sigma/
n^{1/2}$, where $\sigma$ is the standard deviation about the annular
mean, and $n$ is the number of pixels in the annuli.  The solid line
is the best fitting model, and the dotted line is the best fitting
unbroadened PSF.  The points marked with small symbols were not used
in the fits.  These particular sources are illustrated because:
NGC5253-12 is the most diffuse source measured (in angular terms);
NGC5253-43 illustrates a source selected by eye as a probable point
source; NGC1705-4 was selected by eye for its apparent resolution;
Tol1924-416-1 suffers the most central non-linearity/saturation other
than NGC1705-1; NGC4670-7 is a typical barely resolved cluster; and
NGC3690-1 is a typical unresolved SSC.

{Fig.~{14}} -- The main panel plots the effected radius,
$R_e$, of cluster candidates against absolute magnitude, $M_{220}$.
The vertical line shows the confusion limit with individual stars.
The dotted line has a constant surface brightness ($\mu_{220,0} =
9.57~{\rm mag~arcsec^{-2}}$) 625 times more intense than the average
integrated $\mu_{e,0}$ of starbursts. Typical external error bars are
shown in the upper left.  The correspondence between symbols is shown
in the lower panel.  The data points for the SSCs in NGC~1569 and
NGC1705-1 were measured from planetary camera data as discussed in the
text.  The $M_{220}$ of the NGC~1569 clusters were derived from their
$M_{555}$ (\markcite{O'Connell et al., 1994}) assuming they have the
same $M_{220} - M_{555}$ as NGC1705-1.  The right panel shows the
histogram of $R_e$ of Galactic globular clusters from \markcite{van
den Bergh
et al.\ (1991)}.

{Fig.~{15}} -- Luminosity function of the lightly
reddened SSCs.  The smooth dotted line is a luminosity function of the
form $\phi(L) \propto L^{-2}$.  The normalization is to the total
number (59) of objects with $-15 \geq \Muv > -18$.

{Fig.~{16}} -- Evolution of various properties in the
F220W passband.  The panels on the left are for ISB models, those on
the right are for CSF models. In all panels solid lines represent the
stellar component of the LH models, the dotted lines represent the
total light (stellar plus nebular) from the LH models, and the dashed
lines the GISSEL models (in which only the stellar component is
modeled).  The dash dot dash line at $M_{220} = -10.7$ indicates the
peak absolute magnitude of a star having $m_* = 100 \Msun$.

{Fig.~{17}} -- Effect of changing $m_u$ while keeping the
mass normalization constant. The curve shows the difference in
absolute magnitude, \Muv, between two CSF models at an age of 10
Myr. Both have a Salpeter (1955) and the same mass in stars in the
range $1 \leq m_*/\Msun \leq 3$.  One has a fixed IMF upper limit of
$m_u = 100$~\Msun\ while in the other $m_u$ varies between 3 and 120
\Msun.

{Fig.~{18}} -- The ratio of total to selective extinction
$X_{220}$ as a function of \ebv\ in the top panel and $E(\beta)$ in
the bottom panel.  The top panel shows $X_{220}$ for the four
reddening laws and the three CSF models of LH, as well as the
polynomial fit to $X_{220}$ for each law. In the bottom panel only the
transformed fits are shown.

\clearpage

\pagestyle{empty}

\pino

\clearpage

\begin{planotable}{l r r r l}
\tablecaption{Properties of the sample galaxies\label{t:prop}}
\tablewidth{17cm}
\tablehead{\colhead{Galaxy} & \colhead{$V_r$} & \colhead{$D$} &
\colhead{\small $M_B$} & \colhead{Morphology}\\
\colhead{(1)} & \colhead{(2)} & \colhead{(3)} & \colhead{(4)} &
\colhead{(5)} }
\startdata
NGC~1705    &  629 &  6.2    & --16.4 & Amorphous/BCD \nl
IZw18       &  756 & 14.3    & --14.7 & BCD           \nl
NGC~3310    &  980 & 17.9    & --20.1 & SAB(r)bc pec   \nl
NGC~3690    & 2992 & 44.\zsp & --21.3 & SBm? pec      \nl
NGC~3991    & 3192 & 46.\zsp & --19.9 & Im            \nl
NGC~4670    & 1069 & 14.6    & --17.8 & SB(s)0/a pec: \nl
NGC~5253    &  404 &  4.1    & --17.4 & Amorphous/BCD \nl
Tol1924-416 & 2843 & 37.\zsp & --19.9 & Blue compact  \nl
NGC~7552    & 1585 & 19.6    & --20.2 & (R')SB(s)ab
\tablecomments{Keys to columns 2 to 8 follow. \\
Col.2~-- Heliocentric radial velocity in \kms.  The source for the
measurements is the RC3 (de Vaucouleurs et al., 1991) except for
NGC~1705, MFDC; NGC~3690, Mazarella
et al.\ (1993); Tol1924-416, Iye et al.\ (1987). \\
Col.3~-- Distance in Mpc derived from $V_r$ using the linear
Virgo-centric flow model of Schechter (1980) with parameters $\gamma =
2$, $V_{\rm Virgo} = 976$ \kms, $w_\odot = 220$ \kms\ (Binggeli et al.,
1987), and $D_{\rm Virgo} = 15.9$ Mpc (i.e.\ $H_0 = 75$ \kms\,
Mpc$^{-1}$), except for NGC~5253 for which the direct Cepheid $D$ of
Sandage et al.\ (1994) is adopted.  The solution for NGC~4670
is triple valued, we have chosen the middle value (the other
solutions are $D = 11.0, 21.7$ Mpc). \\
Col.4~-- {\em B\/} band absolute magnitude corrected for only Galactic
extinction; i.e.\ $M_B = m_B - 5\log(D/1~{\rm Mpc}) - 25 - A_{\rm B,gal}$.
Both the apparent magnitudes $m_B$, and the Galactic extinction
$A_{\rm B,gal}$ (tabulated in table 2) were extracted from the
NED database.  The original sources are the RC3 (de Vaucouleurs et al.,
1991) and Burstein \&\ Heiles (1982, 1984), respectively. \\
Col.5~-- Morphological classification, mostly taken from the NED data
base (as of mid 1993). \\ }
\end{planotable}

\clearpage

\begin{planotable}{l c l c r r r r r }
\tablecaption{Reddening and extinction.\label{t:redex}}
\tablewidth{170mm}
\tablehead{\colhead{Galaxy} & \colhead{\ebvgal} & \colhead{\ebvbd} &
\colhead{\ebvuv} & \colhead{$E(\beta)$} & \colhead{$A_{\rm 220,gal}$} &
\colhead{$A_{\rm 220,T}$} & \colhead{$A_{\rm B,gal}$} &
\colhead{$A_{\rm H\alpha,T}$} \nl
\colhead{(1)} & \colhead{(2)} & \colhead{(3)} & \colhead{(4)} & \colhead{(5)} &
\colhead{(6)} & \colhead{(7)} & \colhead{(8)} & \colhead{(9)}}
\startdata
NGC~1705    & ~~~0.044~~~ & 0.00 (1)   & ~~~0.003~~~ & 0.02 & 0.37 & 0.40
& 0.18 & 0.11 \nl
IZw18       & 0.005 & 0.09 (2)   & 0.004 & 0.03 & 0.04 & 0.08 & 0.02 & 0.24 \nl
NGC~3310    & 0.000 & 0.23 (3)   & 0.177 & 1.43 & 0.00 & 1.52 & 0.00 & 0.55 \nl
NGC~3690    & 0.000 & 0.66 (4)   & 0.218 & 1.76 & 0.00 & 1.86 & 0.00 & 1.58 \nl
NGC~3991    & 0.005 & 0.08 (5,6) & 0.071 & 0.57 & 0.04 & 0.65 & 0.02 & 0.22 \nl
NGC~4670    & 0.010 & 0.22 (7)   & 0.106 & 0.85 & 0.08 & 0.99 & 0.04 & 0.53 \nl
NGC~5253    & 0.044 & 0.00 (1)   & 0.134 & 1.08 & 0.37 & 1.52 & 0.18 & 0.13 \nl
Tol1924-416 & 0.073 & 0.02 (1)   & 0.032 & 0.26 & 0.62 & 0.90 & 0.30 & 0.23 \nl
NGC~7552    & 0.000 & 0.70 (1)   & 0.369 & 2.98 & 0.00 & 3.13 & 0.00 & 1.68 \nl

\tablecomments{$E$ denotes color excess (reddening) in
magnitudes (except for $E(\beta)$), and $A$ values are extinctions
also in magnitudes.  A ``gal'' subscript denotes reddening or
extinction arising in our Galaxy, while a ``T'' subscript denotes
total = Galactic + internal.  Keys to columns 2 to 9 follow:\\
Col.2~-- Galactic color excess $\ebvgal = A_{\rm B,gal} / 4.1$, where
$A_{\rm B,gal}$ is the {\em B\/} band galactic extinction listed in
col.8.  \\
Col.3~-- Internal reddening (ie total - Galactic) determined from
published Balmer Decrement measurements and assuming case B
recombination.  The numbers in parenthesis are the reference to the
measurements: 1.\ C94; 2.\ Skillman \&\ Kennicutt (1993); 3.\
Pastoriza et al.\ (1993); 4.\ Mazarella \&\ Boroson (1993); 5.\
Osterbrock \&\ Shaw (1988); 6.\ Keel et al.\ (1985); 7.\ Hunter et al.\
(1994b). \\
Col.4~-- Internal reddening of $B-V$ determined from $\beta$ color
excess (see col.5) calculated using the formulism in the appendix and
the K94 extinction law. \\
Col.5~-- Internal color excess in $\beta$: $E(\beta) = \beta + 2.5$,
where the $\beta$ values are those in col.6 of table 4, which have
been corrected for galactic extinction. \\
Col.6~-- Foreground extinction in the F220W bandpass derived from
\ebvgal\ using the formulism in the appendix for the Seaton (1979)
extinction law. \\
Col.7~-- Total extinction in the F220W bandpass where the internal
component is derived from \ebvuv\ using the formulism in the appendix
for the K94 extinction law.\\
Col.8~-- Foreground {\em B\/} band Galactic extinction, $A_{\rm
B,gal}$, from Burstein \&\ Heiles (1984).  \\
Col.9~-- Total extinction at \Halpha.  The Galactic component is
calculated using the Seaton (1979) extinction law, while the internal
component is calculated from \ebvbd\ using the K94 extinction law.}
\end{planotable}

\clearpage

\begin{planotable}{llrrllr}
\tablecaption{FOC observing log\label{t:obslog}}
\tablewidth{0pc}
\tablehead{\colhead{Galaxy} & \colhead{Rootname} & \colhead{R.A.} &
\colhead{Dec.} & \colhead{Filters} & \colhead{UT date} &
\colhead{$\Delta t$} \nl
 \colhead{ } & \colhead{ } & \colhead{(coordinate} &
\colhead{center J2000)} & \colhead{ } & \colhead{(d/m/y h:m)}
& \colhead{($s$)}}
\startdata
NGC~1705a    & x19p5101t & 04 54 13.38 & --53 21 38.5 &
F220W,F1ND & 28/02/93 08:28 & 497 \nl
NGC~1705b    & x19p0101t & 04 54 13.41 & --53 21 38.9 &
F220W,F1ND & 11/04/93 23:35 & 497 \nl
IZw18        & x19p0201t & 09 34 01.91 &  +55 14 27.8 &
F220W      & 11/03/93 11:36 & 1497 \nl
NGC~3310     & x19p0301t & 10 38 45.84 &  +53 30 12.5 &
F220W      & 17/02/93 00:00 &  197 \nl
NGC~3690     & x19p0401t & 11 28 30.90 &  +58 33 45.1 &
F220W      & 11/04/93 10:27 &  897 \nl
NGC~3991     & x19p0501t & 11 57 31.77 &  +32 20 31.0 &
F220W      & 16/02/93 17:27 &  397 \nl
NGC~4670     & x19p0601t & 12 45 17.27 &  +27 07 32.3 &
F220W      & 23/05/93 05:14 &  297 \nl
NGC~5253     & x19p0701m & 13 39 55.99 & --31 38 26.6 &
F220W,F1ND & 21/02/93 22:56 &  497 \nl
Tol1924--416 & x19p0801t & 19 27 58.35 & --41 34 30.0 &
F220W      & 11/04/93 09:14 &  447 \nl
NGC~7552     & x19p0901t & 23 16 10.85 & --42 35 03.0 &
F220W      & 07/05/93 23:04 &  997 \nl
\end{planotable}


\begin{planotable}{l c c r r r}
\tablecaption{Comparison with IUE and FOS properties \label{t:spprop}}
\tablewidth{17cm}
\tablehead{\colhead{Galaxy} & \colhead{$\muv({\rm FOC})$} &
\colhead{$\muv({\rm IUE})$} & \colhead{$\beta_{\rm raw}({\rm IUE})$} &
\colhead{$\beta_{\rm raw}({\rm FOS})$} & \colhead{$\beta$} \nl
\colhead{(1)} & \colhead{(2)}& \colhead{(3)}& \colhead{(4)}& \colhead{(5)}&
\colhead{(6)}}
\startdata
NGC~1705    & $11.86 \pm 0.04$ & $11.98 \pm 0.05$ & --2.40 &        & --2.48\nl
IZw18       & $14.49 \pm 0.05$ & $14.67 \pm 0.15$ & --2.46 &        & --2.48\nl
NGC~3310    & $12.08 \pm 0.03$ & $11.97 \pm 0.18$ & --1.07 &        & --1.07\nl
NGC~3690    & $12.97 \pm 0.03$ & $13.69 \pm 0.18$ & --1.40 & --0.74 & --0.74\nl
NGC~3991    & $13.13 \pm 0.08$ & $13.06 \pm 0.10$ & --1.92 &        & --1.94\nl
NGC~4670    & $12.57 \pm 0.05$ & $12.72 \pm 0.06$ & --1.63 & --1.83 & --1.63\nl
NGC~5253    & $11.75 \pm 0.04$ & $11.69 \pm 0.08$ & --1.34 &        & --1.43\nl
Tol1924-416 & $13.25 \pm 0.07$ & $13.49 \pm 0.08$ & --2.11 &        & --2.24\nl
NGC~7552    & $13.14 \pm 0.05$ & $13.12 \pm 0.19$ &   0.48 & --0.22 &   0.48\nl
\tablecomments{Keys to columns 2 to 6 follow. \\
Col.2 -- The apparent magnitude, $\muv$, extracted from the FOC images
with a circular aperture having $R = 6.94''$ yielding the same area as
the IUE extraction aperture of K93.  The error combines the photon
statistics error in the total number of counts with the uncertainty
induced by changing the background level by $\pm 1\sigma$. \\
Col.3 -- \muv\ derived from the IUE spectra of K93.  The error was
derived by interpolating the errors reported in Table 8 of K93.\\
Col.4 -- UV spectral slope from the IUE spectra.  No corrections for
reddening (either Galactic foreground or internal) have been
made. Typical formal errors in $\beta$ are 0.02.  However, $\beta$ is
sensitive to the exact placement of the continuum windows, so more
realistic errors are $\sim 0.15$ (Robert et al., 1995). \\
Col.5 -- Same as col.4 but from the FOS spectra of Robert et al.\
(1995) which were obtained with a 1\as\ diameter aperture.\\
Col.6 -- Adopted UV slope after correction for foreground Galactic
extinction (see Table 2).}
\end{planotable}

\clearpage

\begin{planotable}{l r r r r r r r r r}
\tablecaption{Surface Photometry Extraction Parameters \label{t:ellpar}}
\tablewidth{0pc}
\tablehead{\colhead{Name} & \colhead{$\Delta \alpha_i$} &
\colhead{$\Delta \delta_i$} & \colhead{$a_i$} & \colhead{$\phi_i$} &
\colhead{$\Delta \alpha_o$} & \colhead{$\Delta \delta_o$} &
\colhead{$a_o$} & \colhead{$(a/b)_o$} & \colhead{$\phi_o$} \nl
 & \colhead{(\as)} & \colhead{(\as)} & \colhead{(\as)} & \colhead{(\dg)} &
\colhead{(\as)} & \colhead{(\as)} & \colhead{(\as)} &  & \colhead{(\dg)} }
\startdata
NGC1705     &   0.000 &   0.000 & 2.00 &   30.00 & --1.380 & --1.170 &
7.50 & 1.300 &   72.70   \nl
IZw18       & --0.394 & --0.313 & 1.00 & --37.43 & --1.277 & --1.854 &
6.70 & 1.964 & --37.43   \nl
NGC3310     &   0.000 &   0.000 & 1.50 & --52.00 & --0.640 & --1.270 &
4.50 & 1.125 & --52.00   \nl
NGC3690-BC  &   1.687 &   3.157 & 1.00 & --12.10 &   1.112 &   0.778 &
5.20 & 1.480 & --12.10   \nl
NGC3690-Ab  &   7.423 & --4.611 & 1.00 &    0.00 &   7.423 & --4.611 &
4.20 & 1.000 &    0.00   \nl
NGC3690-Aa  &  13.292 & --3.670 & 1.00 &    0.00 &  13.292 & --3.670 &
4.20 & 1.000 &    0.00   \nl
NGC3991     &   0.000 &   0.000 & 1.00 &   14.13 &   1.322 & --3.451 &
7.30 & 2.690 &   14.13   \nl
NGC4670     &   0.000 &   0.000 & 1.00 &   75.60 &   2.133 & --0.967 &
4.30 & 1.830 &   75.60   \nl
NGC5253     &   0.000 &   0.000 & 2.50 &   22.00 &   0.000 &   0.000 &
8.60 & 1.800 &   22.00   \nl
Tol1924-416 &   0.608 & --0.090 & 1.00 &   78.70 &   0.712 & --0.344 &
4.90 & 1.870 &   93.60   \nl
NGC7552     &   0.500 & --0.275 & 1.00 &   83.70 &   1.459 & --1.357 &
7.10 & 1.500 &   83.70   \nl
\end{planotable}


\begin{planotable}{l r c c c c }
\tablecaption{Surface Photometry Results \label{t:surfres}}
\tablewidth{0pc}
\tablehead{\colhead{Name} & \colhead{$a_{\rm max}$} & \colhead{$a_e$}
& \colhead{$m_T$} & \colhead{$\mu_e$} & \colhead{$f_{\rm clumpy}$}\nl
 & \colhead{(\as)} & \colhead{(\as)} & \colhead{(mag)} &
\colhead{(mag/arcsec$^2$)}}
\startdata
NGC 1705 (total)        & 11.20 & 0.85 & 11.77 & 14.0: & 0.52 \nl
NGC 1705 (no NGC1705-1) & 11.20 & 2.75 & 12.36 & 16.51 & 0.17 \nl
IZw18                   &  8.75 & 1.95 & 14.45 & ~~~17.74~~~ & 0.05 \nl
NGC 3310                & 12.60 & 7.75 & 11.08 & 17.39 & 0.07 \nl
NGC 3690-BC             & 13.30 & 3.20 & 12.95 & 17.23 & 0.23 \nl
NGC 3690-Ab             &  4.20 & 1.25 & 15.15 & 17.63 & 0.21 \nl
NGC 3690-Aa             &  2.80 & 1.20 & 15.47 & 17.86 & 0.31 \nl
NGC 3690 (sum)          &       &      & 12.73 &       & 0.24 \nl
NGC 3991                & 16.45 & 4.00 & 12.86 & 17.22 & 0.14 \nl
NGC 4670                & 14.00 & 3.05 & 12.46 & 16.43 & 0.18 \nl
NGC 5253                & 15.40 & 6.10 & 11.43 & 16.93 & 0.21 \nl
TOL1924-416             &  9.80 & 2.70 & 13.22 & 17.03 & 0.16 \nl
NGC 7552                &  8.40 & 1.45 & 13.12 & 15.88 & 0.21 \nl
\end{planotable}

\clearpage

\begin{planotable}{l c c c c}
\tablecaption{Compact source cross-identifications.\label{t:crossid}}
\tablewidth{0pc}
\tablehead{\colhead{Our name} & \colhead{Other name} & \colhead{Ref.}&
\colhead{Other name} & \colhead{Ref.}}
\startdata
NGC5253-1     & NK1     & CP89 & & \nl
NGC5253-12    & NK2     & CP89 & & \nl
NGC5253-2     & NK3     & CP89 & & \nl
NGC5253-4,7   & NK4     & CP89 & & \nl
NGC1705-1     & ~~~nucleus~~~ & MFDC & A & MMT \nl
NGC1705-4     & j       & MFDC & & \nl
NGC3310-5,9   & A       & P93  & ~~~Jumbo~~~ & BH81 \nl
NGC3310-3,8   & C       & P93  & & \nl
NGC3310-2,6   & E       & P93  & & \nl
NGC3690-5,8   & B1      & WW91 & & \nl
NGC3690-2     & B2      & WW91 & Ac & MB93 \nl
NGC3690-4 (Ab) &  S     & F87  & Ab & MB93 \nl
NGC3690-3 (Aa) &  W     & F87  & Aa & MB93 \nl
\tablerefs{CP89: Caldwell \&\ Phillips, 1989; MFDC: Meurer
et al.\ (1992); MMT: Melnick et al.\ (1985); P93:
Pastoriza et al., 1993; BH81: Balick \&\ Heckman, 1981; WW91:
Wynn-Williams et al.\
(1991); MB93: Mazarella \&\ Boroson, 1993; F87: Friedman et al., 1987.}
\end{planotable}

\clearpage

\begin{planotable}{l r r c r c r r}
\small
\tablecaption{Fluxes\label{t:flux}}
\tablewidth{17cm}
\tablehead{\colhead{Galaxy} & \colhead{\em FIR} & \colhead{$F_{220}$} &
\colhead{$F_{220,0}$} &
\colhead{\Foha} & \colhead{Ref.} & \colhead{IRX} &
\colhead{$\Foha/F_{220,0}$}\nl
\colhead{(1)} & \colhead{(2)} & \colhead{(3)} & \colhead{(4)} &
\colhead{(5)} & \colhead{(6)} & \colhead{(7)} & \colhead{(8)}}
\startdata
NGC~1705    & 51.2    & 235. & 240.   & 3.51  & 1   & --0.66   &  0.0146  \nl
IZw18       & $< 3.2$ & 15.1 & 15.6   & 0.53  & 2   & $<-0.67$ &  0.0336  \nl
NGC~3310    & 1490    & 311. & 1260   & 16.0  & 3,4 &   0.68   &  0.0128  \nl
NGC~3690    & 2840    & 68.2 & 666.   & 6.75  & 5,6 &   1.62   &  0.0101  \nl
NGC~3991    & 89      & 65.4 & 58.5,115 & 0.326 & 7 &   0.13   &  0.0056  \nl
NGC~4670    & 142     & 87.4 & 202.   & 3.42  & 8   &   0.21   &  0.0169  \nl
NGC~5253    & 1360    & 334. & 960.   & 32.5  & 9   &   0.61   &  0.0339  \nl
Tol1924-416 & 67.7    & 74.9 & 64.4,96.6 & 0.703 & 10 & --0.04 &  0.0110  \nl
NGC~7552    & 3620    & 47.6 & 849.   & 3.29  & 11  &   1.88   &  0.0039  \nl
\tablecomments{Fluxes are in units of $10^{-12}$ \FU. Keys to columns
2 to 6 follow. \\
Col.2 -- The infrared flux {\em FIR\/} is derived
from the IRAS Faint Source Catalog (Moshir et al., 1990) measurements
in the 60\micron\ and 100\micron\ bands using the definition of Helou
et al.\ (1988, in their appendix).  IZw18 was not detected by IRAS so
is given as an upper limit.  NGC~3991 was not detected in the
100\micron\ band; we assumed a 100\micron\ flux of half the upper
limit reported in the IRAS catalog. For NGC~3690, we assume that its
flux is 60\%\ of {\em FIR\/} for the Arp299 system, following Gehrz
et al.\ (1983)\\
Col.3 -- $F_{220}$ is the {\em total} UV flux ($\lambda f_\lambda$ as
defined in appendix), derived from $m_T$ in Table 6, corrected for
Galactic extinction using the $A_{\rm 220,gal}$ values in Table 2. \\
Col.4 -- $F_{220,0}$ is the UV flux corrected for Galactic and
internal extinction using the $A_{\rm 220,T}$ values in Table 2.  The
first flux reported for NGC~3991 and Tol1924-416 is taken for a
circular aperture matching in area that of the \Halpha\ observations
as noted below, the second flux is the total UV flux. It is the former
value that is used in calculating $\Foha/F_{220,0}$.\\
Col.5 -- \Foha\ is the \Halpha\ flux corrected for extinction using
the $A_{\rm H\alpha,T}$ value listed in Table 2.  The fluxes were
taken from the sources listed in col.\ 6 and measured through large
enough of an aperture to recover most of the total flux with the
exception of the NGC~3991 and Tol1924-416 data where a 4.7\as\
diameter circular aperture and a 4\as\ wide square aperture,
respectively, were employed. Note that the IZw18 measurement is for
region ``A'' of Dufour \&\ Hester (1990), the only portion imaged in
this study, and that the NGC~7552 measurement is for the ring only. \\
Col.6 -- Reference for \Foha\ measurements.  1.\ MFDC; 2.\ Dufour and
Hester (1990); 3.\ Kennicutt \&\ Kent (1983); 4.\ Pastoriza et al.\
(1993); 5.\ Armus et al.\ (1990); 6.\ Friedman et al.\ (1987); 7.\ Keel
et al.\ (1985); 8.\ Marlowe et al.\ (1995); 9.\ Walsh \&\ Roy (1989);
10.\ Bergvall (1985); 11.\ Forbes et al.\ (1994a).  The first of the
two references for NGC~3310 and NGC~3690 provided a \Halpha\ +
\fion{N}{II} flux, and the second the $\fion{N}{II}/\Halpha$ ratio
necessary to convert to a pure \Halpha\ flux.
}
\end{planotable}

\clearpage

\begin{planotable}{l r r c r r l r r}
\tablecaption{Intrinsic properties \label{t:intrinsic}}
\tablewidth{170mm}
\tablehead{\colhead{Name} & \colhead{$R_e$} &
\colhead{$R_{\rm max}$} & \colhead{$\mu_{e,0}$} & \colhead{$L_{220}$} &
\colhead{$M_{M*}$} &
\colhead{MSFR} & \colhead{MSFA$_e$} & \colhead{SFA$_e$} \nl
\colhead{(1)} & \colhead{(2)} & \colhead{(3)} & \colhead{(4)} &
\colhead{(5)} & \colhead{(6)} & \colhead{(7)} & \colhead{(8)} & \colhead{(9)}}
\startdata
NGC 1705 (total)        & 25.6 &  295 & 13.60 &  1.09 & 0.47 &
0.040 & 9.4  & 52 \nl
NGC 1705 (-- NGC1705-1) & 81.0 &  295 & 16.11 & 0.634 & 0.27 &
0.024 & 0.54 & 3.0 \nl
IZw18                   &  126 &  433 & 17.66 & 0.367 & 0.16 &
0.014 & 0.13 & 0.71 \nl
NGC 3310                &  634 & 1093 & 15.87 &  47.9 & 21.  &
1.79  & 0.67 & 3.7 \nl
NGC 3690-BC             &  610 & 2332 & 15.37 &  70.8 & 31.  &
2.7   & 1.1  & 5.9 \nl
NGC 3690-Ab             &  267 &  896 & 15.77 &  9.33 & 4.1  &
0.35  & 0.74 & 4.1 \nl
NGC 3690-Aa             &  256 &  597 & 16.00 &  7.08 & 3.0  &
0.26  & 0.60 & 3.3 \nl
NGC 3690 (sum)          &      &      &       &  87.7 & 38.  &
3.2   &      &     \nl
NGC 3991                &  664 & 2237 & 16.57 &  27.9 & 12.  &
1.0   & 0.36 & 2.0 \nl
NGC 4670                &  175 &  733 & 15.44 &  5.50 & 2.4  &
0.21  & 1.0  & 5.5 \nl
NGC 5253                & 99.9 &  228 & 15.41 &  1.84 & 0.8  &
0.068 & 1.0  & 5.7 \nl
Tol1924-416             &  415 & 1292 & 16.13 &  16.2 & 7.1  &
0.61  & 0.53 & 2.9 \nl
NGC 7552                &  135 &  652 & 12.75 &  39.9 & 17.  &
1.45  & 12.  & 66. \nl
\tablecomments{Keys to columns 2 to 8 follow:\\
Col.2~-- Effective radius in pc. \\
Col.3~-- Maximum or total radius (see text) in pc. \\
Col.4~-- Extinction corrected mean surface brightness within $R_e$ in
units of mag arcsec$^{-2}$.  The $A_{\rm 220,T}$ values
listed in Table 2 were used for the extinction correction. \\
Col.5~-- UV Luminosity in the F220W band: $L_{220} = 4\pi D^{2}
F_{220,0}$, in units of $10^{42}\, {\rm erg\, s^{-1}}$, where
$F_{220,0}$ is the total UV extinction corrected flux listed in Table
8. \\
Col.6~-- Mass in stars massive stars, that is those with $5 \leq
m_*/\Msun \leq 100$ assuming a Salpeter (1955) calculated from
$L_{220}$ as explained in the text.  Units are $10^6 \Msun$.\\
Col.7~-- Massive star formation rate, that is the rate of formation of
stars with $5 \leq m_*/\Msun \leq 100$ assuming a constant star
formation rate.  Units are $\Msun\ {\rm yr^{-1}}$.   \\
Col.8~-- Mean massive star formation rate per area within $R_e$ in units of
$\Msun\, {\rm Kpc^{-2}\, yr^{-1}}$.
Col.9~-- Star formation rate per area within $R_e$ for stars with $0.1
\leq m_*/\Msun \leq 100$, in units of
$\Msun\, {\rm Kpc^{-2}\, yr^{-1}}$. }
\end{planotable}

\clearpage

\begin{planotable}{lrrcrrl}
\small
\tablecaption{Compact sources\label{t:sources}}
\tablewidth{150mm}
\tablehead{\colhead{Object} & \colhead{$\Delta\alpha$} &
\colhead{$\Delta\delta$} & \colhead{\muv} & \colhead{\Muv} & $R_e$ & Notes \nl
 & \colhead{(\as)} & \colhead{(\as)} & \colhead{(mag)} &
\colhead{(mag)} & \colhead{(pc)} & \nl
\colhead{(1)} & \colhead{(2)} & \colhead{(3)} & \colhead{(4)} &
\colhead{(5)} & \colhead{(6)} & \colhead{(7)} }
\startdata
NGC5253-1      & $  0.00$ & $  0.00$ & $17.56\pm 0.14$ &
$-13.8$ &   2.0      & a,c,d \nl
NGC5253-2      & $  0.76$ & $ -4.34$ & $17.07\pm 0.06$ &
$-13.4$ &   1.5      & c \nl
NGC5253-3      & $ -6.95$ & $ -6.58$ & $18.08\pm 0.11$ &
$-12.7$ &   1.8      & c \nl
NGC5253-4      & $ -5.01$ & $ -2.25$ & $17.59\pm 0.06$ &
$-12.5$ &   0.9      &  \nl
NGC5253-5      & $  0.67$ & $ -0.41$ & $17.31\pm 0.05$ &
$-12.5$ & $\leq 0.7$ & d \nl
NGC5253-6      & $  5.46$ & $  6.36$ & $17.54\pm 0.06$ &
$-12.4$ &   0.8      & b \nl
NGC5253-7      & $ -4.36$ & $ -1.81$ & $18.61\pm 0.11$ &
$-12.3$ &   1.8      &  \nl
NGC5253-8      & $ -2.89$ & $  2.41$ & $18.40\pm 0.11$ &
$-12.1$ &   1.3      &  \nl
NGC5253-9      & $  6.49$ & $ -0.45$ & $17.96\pm 0.06$ &
$-12.1$ &   1.2      &  \nl
NGC5253-10     & $ -0.19$ & $  0.64$ & $17.56\pm 0.06$ &
$-12.1$ & $\leq 0.7$ & d \nl
NGC5253-11     & $ -0.62$ & $ -6.03$ & $18.00\pm 0.08$ &
$-12.1$ &   1.0      &  \nl
NGC5253-12     & $  0.91$ & $  2.52$ &  \nodata        &
$-12.1$ &   2.9      & o \nl
NGC5253-16     & $  8.06$ & $  4.84$ & $18.18\pm 0.07$ &
$-11.6$ & $\leq 0.7$ &  \nl
NGC5253-26     & $  0.48$ & $  7.98$ & $18.52\pm 0.08$ &
$-11.2$ & $\leq 0.7$ & e \nl
NGC5253-43     & $ -4.09$ & $  3.78$ & $18.70\pm 0.09$ &
$-10.9$ & $\leq 0.7$ & e \nl
NGC5253-93     & $  0.79$ & $ 10.71$ & $19.49\pm 0.13$ &
$-10.2$ & $\leq 0.7$ & e \nl
NGC1705-1      & $  0.00$ & $  0.00$ & $12.72\pm 0.00$ &
$-16.6$ &   1.1      & a,h \nl
NGC1705-2      & $ -0.60$ & $ -0.72$ & $16.05\pm 0.03$ &
$-13.6$ &   1.2      &  \nl
NGC1705-3      & $  0.79$ & $  1.31$ & $18.49\pm 0.08$ &
$-11.4$ &   1.4      & f \nl
NGC1705-4      & $  0.96$ & $  3.29$ & $19.19\pm 0.10$ &
$-10.9$ &   1.8      &  \nl
NGC1705-6      & $ -1.06$ & $ -0.46$ & $18.31\pm 0.12$ &
$-10.7$ & $\leq 1.0$ & f \nl
NGC1705-8      & $ -3.15$ & $ -5.30$ & $19.12\pm 0.09$ &
$-10.5$ &   1.4      & b \nl
NGC1705-9      & $  1.30$ & $ -2.72$ & $19.11\pm 0.09$ &
$-10.4$ & $\leq 1.0$ & d \nl
NGC1705-12     & $  4.36$ & $ -1.82$ & $19.07\pm 0.08$ &
$-10.2$ & $\leq 1.0$ & e \nl
NGC1705-15     & $ -6.55$ & $ -6.14$ & $19.34\pm 0.09$ &
$-10.2$ &   1.1      &  \nl
NGC1705-17     & $ -4.15$ & $ -5.04$ & $19.22\pm 0.09$ &
$-10.2$ & \nodata    & g \nl
NGC1705-25     & $ -1.88$ & $  3.38$ & $19.28\pm 0.09$ &
$ -9.9$ & $\leq 1.0$ & g \nl
NGC1705-46     & $  1.87$ & $  5.59$ & $19.81\pm 0.11$ &
$ -9.4$ & $\leq 1.0$ & e \nl
IZw18-1        & $  0.00$ & $  0.00$ & $19.50\pm 0.07$ &
$-11.8$ &   3.2      & a \nl
IZw18-2        & $  0.78$ & $ -0.86$ & $20.22\pm 0.12$ &
$-11.4$ &   4.2      & d \nl
IZw18-3        & $  1.22$ & $ -0.39$ & $20.11\pm 0.10$ &
$-11.3$ &   4.1      & d \nl
IZw18-4        & $  3.82$ & $ -4.75$ & $20.70\pm 0.11$ &
$-10.8$ &   3.7      & c \nl
IZw18-5        & $  0.84$ & $ -0.48$ & $20.06\pm 0.09$ &
$-10.8$ & $\leq 2.4$ & d \nl
IZw18-6        & $  1.85$ & $ -0.64$ & $20.78\pm 0.11$ &
$-10.3$ &   2.5      &  \nl
IZw18-10       & $  0.28$ & $ -0.50$ & $20.66\pm 0.13$ &
$-10.1$ & $\leq 2.4$ &  \nl
IZw18-12       & $  0.58$ & $ -1.05$ & $20.59\pm 0.12$ &
$ -9.9$ & $\leq 2.4$ & d \nl
NGC4670-1      & $  0.00$ & $  0.00$ & $16.57\pm 0.04$ &
$-15.1$ & $\leq 2.4$ & a,b \nl
NGC4670-2      & $  0.11$ & $ -0.38$ & $17.20\pm 0.08$ &
$-14.8$ &   3.4      & b \nl
NGC4670-3      & $ -1.92$ & $ -1.91$ & $17.02\pm 0.04$ &
$-14.8$ & $\leq 2.4$ & d \nl
NGC4670-4      & $ -2.01$ & $ -1.16$ & $17.29\pm 0.05$ &
$-14.5$ & $\leq 2.4$ & b \nl
NGC4670-5      & $ -5.13$ & $ -1.21$ & $17.38\pm 0.05$ &
$-14.4$ & $\leq 2.4$ & c \tablebreak
NGC4670-6      & $ -1.29$ & $ -1.61$ & $17.44\pm 0.05$ &
$-14.4$ & $\leq 2.4$ &  \nl
NGC4670-7      & $ -0.29$ & $  0.35$ & $17.95\pm 0.08$ &
$-14.3$ &   3.5      &  \nl
NGC4670-8      & $ -2.10$ & $ -2.17$ & $17.68\pm 0.06$ &
$-14.1$ &   3.1      & b \nl
NGC4670-9      & $ -5.06$ & $ -1.63$ & $17.76\pm 0.05$ &
$-14.1$ & $\leq 2.4$ &  \nl
NGC4670-10     & $ -2.24$ & $ -2.51$ & $18.17\pm 0.07$ &
$-13.8$ &   2.6      &  \nl
NGC4670-21     & $-10.74$ & $ -1.76$ & $19.37\pm 0.10$ &
$-12.4$ & \nodata    &  \nl
NGC3310-1      & $  0.00$ & $  0.00$ & $16.13\pm 0.03$ &
$-16.6$ & $\leq 3.0$ & d \nl
NGC3310-2      & $ -3.38$ & $ 11.01$ & $17.28\pm 0.05$ &
$-15.7$ & $\leq 3.0$ & c \nl
NGC3310-3      & $  0.10$ & $  7.35$ & $17.46\pm 0.05$ &
$-15.4$ & $\leq 3.0$ &  \nl
NGC3310-4      & $ -2.13$ & $ -1.14$ & $17.69\pm 0.06$ &
$-15.1$ & $\leq 3.0$ &  \nl
NGC3310-5      & $-15.02$ & $ -0.96$ & $18.39\pm 0.10$ &
$-14.7$ &   3.2      &  \nl
NGC3310-6      & $ -2.98$ & $ 10.59$ & $18.26\pm 0.08$ &
$-14.6$ &   3.2      &  \nl
NGC3310-7      & $ -9.41$ & $  5.53$ & $18.43\pm 0.09$ &
$-14.6$ &   3.7      &  \nl
NGC3310-8      & $ -0.03$ & $  6.83$ & $18.60\pm 0.10$ &
$-14.6$ &   4.1      &  \nl
NGC3310-9      & $-14.24$ & $ -1.50$ & $18.51\pm 0.12$ &
$-14.4$ & $\leq 3.0$ & d \nl
NGC3310-10     & $-11.03$ & $ -4.08$ & $18.76\pm 0.09$ &
$-14.3$ &   3.7      &  \nl
NGC7552-1      & $  0.00$ & $  0.00$ & $16.47\pm 0.02$ &
$-18.1$ & $\leq 3.2$ & a \nl
NGC7552-2      & $  0.48$ & $ -0.66$ & $16.75\pm 0.04$ &
$-17.9$ & $\leq 3.2$ & j \nl
NGC7552-3      & $ -0.29$ & $  0.21$ & $17.14\pm 0.05$ &
$-17.7$ &   3.5      &  \nl
NGC7552-4      & $ -0.84$ & $ -0.68$ & $17.56\pm 0.05$ &
$-17.3$ &   4.0      &  \nl
NGC7552-5      & $ -1.11$ & $ -0.35$ & $18.06\pm 0.08$ &
$-17.0$ &   4.5      &  \nl
NGC7552-6      & $ -4.34$ & $ -2.51$ & $18.49\pm 0.04$ &
$-16.1$ & $\leq 3.2$ &  \nl
NGC7552-13     & $ -2.47$ & $ -2.09$ & $20.21\pm 0.10$ &
$-14.4$ & \nodata    &  \nl
NGC7552-25     & $ -2.81$ & $ -1.63$ & $20.96\pm 0.16$ &
$-13.6$ & \nodata    &  \nl
TOL1924-416-1  & $  0.00$ & $  0.00$ & $16.71\pm 0.04$ &
$-17.3$ & $\leq 6.2$ & a,c,d \nl
TOL1924-416-2  & $-11.72$ & $ -4.81$ & $17.58\pm 0.04$ &
$-16.2$ & $\leq 6.2$ & d,k \nl
TOL1924-416-3  & $  1.36$ & $ -2.61$ & $17.89\pm 0.05$ &
$-15.9$ & $\leq 6.2$ &  \nl
TOL1924-416-4  & $ -0.87$ & $ -0.61$ & $18.05\pm 0.05$ &
$-15.9$ & $\leq 6.2$ & d \nl
TOL1924-416-5  & $ -5.43$ & $ -0.47$ & $18.32\pm 0.06$ &
$-15.8$ &   8.6      &  \nl
TOL1924-416-6  & $ -1.11$ & $ -0.40$ & $18.29\pm 0.06$ &
$-15.7$ &   6.7      & d \nl
TOL1924-416-7  & $ -3.41$ & $ -0.42$ & $19.31\pm 0.10$ &
$-15.6$ &  16.7      &  \nl
TOL1924-416-8  & $ -1.36$ & $ -0.92$ & $18.85\pm 0.08$ &
$-15.0$ & $\leq 6.2$ &  \nl
TOL1924-416-9  & $ -2.57$ & $ -0.04$ & $19.18\pm 0.08$ &
$-14.7$ & $\leq 6.2$ &  \nl
TOL1924-416-10 & $ -0.21$ & $  0.13$ & $19.41\pm 0.17$ &
$-14.3$ & \nodata    &  \nl
TOL1924-416-11 & $  2.40$ & $ -1.68$ & $19.45\pm 0.10$ &
$-14.3$ & \nodata    &  \nl
TOL1924-416-12 & $-11.46$ & $ -4.76$ & $19.57\pm 0.09$ &
$-14.2$ & \nodata    &  \nl
TOL1924-416-13 & $ -1.35$ & $  0.76$ & $19.62\pm 0.12$ &
$-14.1$ & \nodata    &  \nl
TOL1924-416-14 & $ -4.14$ & $ -0.11$ & $19.69\pm 0.14$ &
$-14.0$ & \nodata    &  \nl
NGC3690-1      & $  0.00$ & $  0.00$ & $16.57\pm 0.02$ &
$-18.5$ & $\leq 7.4$ & a \nl
NGC3690-2      & $ -0.49$ & $ -2.02$ & $17.32\pm 0.04$ &
$-18.3$ &  10.9      &  \nl
NGC3690-3      & $-13.29$ & $ -3.68$ & $17.05\pm 0.04$ &
$-18.0$ & \nodata    & i \tablebreak
NGC3690-4      & $ -7.42$ & $ -4.62$ & $17.75\pm 0.04$ &
$-17.7$ &  10.6      & d \nl
NGC3690-5      & $  0.56$ & $ -4.97$ & $17.53\pm 0.04$ &
$-17.6$ & \nodata    & l \nl
NGC3690-6      & $ -2.06$ & $  4.30$ & $18.26\pm 0.06$ &
$-17.3$ &  12.8      & b \nl
NGC3690-7      & $ -2.88$ & $  3.65$ & $18.07\pm 0.04$ &
$-17.3$ & $\leq 7.4$ & j \nl
NGC3690-8      & $  0.51$ & $ -4.84$ & $18.01\pm 0.06$ &
$-17.1$ & \nodata    & l \nl
NGC3690-9      & $ -3.39$ & $  1.98$ & $18.31\pm 0.04$ &
$-16.8$ & $\leq 7.4$ &  \nl
NGC3690-10     & $ -3.16$ & $  1.37$ & $18.81\pm 0.06$ &
$-16.6$ &   7.9      &  \nl
NGC3690-11     & $-13.12$ & $ -5.16$ & $19.08\pm 0.06$ &
$-16.5$ &  11.4      & k \nl
NGC3690-12     & $ -1.88$ & $  2.32$ & $19.01\pm 0.07$ &
$-16.1$ & \nodata    & m \nl
NGC3690-13     & $ -1.91$ & $  2.54$ & $19.22\pm 0.08$ &
$-15.9$ & \nodata    & m \nl
NGC3690-14     & $ -2.92$ & $  2.15$ & $19.26\pm 0.09$ &
$-15.8$ & \nodata    &  \nl
NGC3690-15     & $ -1.36$ & $  4.68$ & $19.35\pm 0.08$ &
$-15.8$ & \nodata    &  \nl
NGC3690-16     & $ -4.55$ & $  1.14$ & $19.49\pm 0.07$ &
$-15.6$ & \nodata    &  \nl
NGC3991-1      & $  0.00$ & $  0.00$ & $17.06\pm 0.05$ &
$-17.5$ &  12.4      & a,b \nl
NGC3991-2      & $ -0.58$ & $ -0.85$ & $19.12\pm 0.15$ &
$-16.2$ &  21.0      & c \nl
NGC3991-3      & $ -1.75$ & $ -6.05$ & $19.02\pm 0.09$ &
$-15.4$ &  10.7      & b \nl
NGC3991-4      & $  0.27$ & $  0.42$ & $19.42\pm 0.14$ &
$-15.3$ &  15.7      &  \nl
NGC3991-5      & $ -1.33$ & $ -2.75$ & $18.80\pm 0.09$ &
$-15.2$ & \nodata    & n \nl
NGC3991-6      & $ -1.22$ & $ -2.54$ & $19.28\pm 0.11$ &
$-14.7$ & \nodata    & n \nl
NGC3991-7      & $ -4.71$ & $ -5.42$ & $19.69\pm 0.11$ &
$-14.7$ &  11.5      &  \nl
NGC3991-8      & $ -1.62$ & $ -0.92$ & $19.95\pm 0.16$ &
$-14.5$ &  15.7      & b \nl
NGC3991-9      & $ -1.10$ & $ -2.45$ & $19.70\pm 0.15$ &
$-14.3$ & \nodata    &  \nl
NGC3991-10     & $ -0.86$ & $ -2.67$ & $19.99\pm 0.16$ &
$-14.3$ &  11.6      & d \tablebreak
\tablecomments{Keys to columns 2 -- 7 follow. \\
Col.2,3 -- The position offsets relative to the coordinates given in
table~3.  These relative positions should be accurate to about 30 mas (P.\
Hodge, private communication, 1994) with the exception of
the images of NGC~1705a, and Tol1924-416.  These exposures were
initiated 1.5 hours after FOC turn on, which is considered to be too
soon for the geometric distortions to fully stabilize.  A comparison
of matched positions in the two NGC~1705 frames yields a dispersion of
80 mas (3.5 pixels) in relative offsets. \\
Col.4 -- The apparent magnitude, derived from the aperture photometry,
with no correction for the size of the object.  The errors are the
random errors due to photon statistics.  An additional quasi-random
uncertainty of $\sim 0.15$ mag is expected due to positional
variations in the photometric performance of the FOC (Meurer, 1995). \\
Col.5 -- Absolute magnitudes derived from the profile fitting where
available, otherwise from the aperture photometry. \\
Col.6 -- Effective or half light radius determined from profile
fitting.  Sources not fitted are indicated with an ellipses (\ldots). \\
Col.7 -- Notes on the individual sources as follow: (a) Coordinate
system origin for the galaxy. (b) Elongated or double in
appearance. (c) Elliptical shape. (d) Neighbor with $10 < r < 14$
pixels. (e) Selected as probable star. (f) Low signal/noise because of
NGC1705-1 subtraction. (g) $\Delta m_{220} > 0.4$ mag between NGC~1705
frames. (h) $R_e$ from PC image.  See text.  (i) At edge of thin
occulting finger. (j) Reseau within fitting area. (k) Near edge of
frame.  (l) Double source NGC3690-5/8. (m) Double source
NGC3690-12/13. (n) Double source NGC3991-5/6. (o) very diffuse, not
found by DAOFIND. Fit out to $r= 15.5$ pixels. }
\end{planotable}

\clearpage

\begin{planotable}{l r r r r r r}
\tablecaption{Average quantities for ionizing populations. \label{t:avgq}}
\tablewidth{0pc}
\tablehead{& & \colhead{ISB} & & & \colhead{CSF} & \nl
\colhead{quantity} & \colhead{avg} & \colhead{min} & \colhead{max} &
\colhead{avg} & \colhead{min} & \colhead{max}}
\startdata
$220-V$                             &  --2.53 &  --2.92 & --0.95 &
  --2.38 &  --2.88 & --1.63 \nl
$RL$                                & --0.003 & --0.008 &  0.002 &
 --0.001 & --0.003 &  0.002 \nl
$NC$                                &    0.14 &    0.00 &   0.31 &
    0.17 &    0.08 &   0.33 \nl
    0.31 &    0.25 &   0.34 \nl
$\beta$                             &  --2.52 &  --2.70 & --2.26 &
  --2.51 &  --2.65 & --2.32 \nl
$\log(L_{220}/L_{\rm T})$           &  --0.48 &  --0.61 & --0.41 &
  --0.50 &  --0.58 & --0.46 \nl
$\log(M/L_{220})~(m_l =5\,\Msun)$   &  --2.96 &  --3.24 & --2.37 &
  --2.78 &  --3.10 & --2.13 \nl
$\log(M/L_{220})~(m_l =1\,\Msun)$   &  --2.62 &  --2.90 & --2.04 &
  --2.46 &  --2.76 & --1.80 \nl
$\log(M/L_{220})~(m_l =0.1\,\Msun)$ &  --2.22 &  --2.50 & --1.63 &
  --2.04 &  --2.35 & --1.39 \nl
\end{planotable}


\begin{planotable}{c c c c r c c}
\tablecaption{Effects of different reddening laws. \label{t:redprop}}
\tablewidth{0pc}
\tablehead{\colhead{law} & \colhead{$r_\beta$} & \colhead{$a_0$} &
\colhead{$a_1$} &
\colhead{$a_2$} & \colhead{$a_3$} & \colhead{rms}}
\startdata
Galactic        & 1.793 & 8.519 & --0.647 & --0.103 &         & 0.014 \nl
LMC             & 6.896 & 8.069 & --0.309 & --0.105 &         & 0.012 \nl
Starburst (K94) & 8.067 & 8.613 & --0.370 &   0.035 & --0.085 & 0.015 \nl
Starburst (C94) & 4.326 & 6.956 & --0.097 & --0.024 &         & 0.006 \nl
\end{planotable}


\begin{references}
\reference Allen, C.W. 1973, Astrophysical Quantities, Third
Edition, (Athlone, London)
\reference Armus, L., Heckman, T.M., \&\ Miley, G.K. 1990, \apj,
364, 471 (AHM)
\reference Arp, H. \&\ Sandage, A. 1985, \aj, 90, 1163
\reference Ashman, K.M., \&\ Zepf, S.E. 1992, \apj, 384, 50
\reference Balick, B., \&\ Heckman, T. 1981, \aap, 96, 271
\reference Balzano, V.A. 1983, \apj, 268, 602
\reference Barth, A.J., Ho, L.C., Filipenko, A.V., \&\ Sargent,
W.L.W., 1995, AJ, (accepted)
\reference Baxter, D.A. 1993, Instrument Science Report FOC-072,
(STScI, Baltimore)
\reference Baxter, D.A., Greenfield, P.E., Hack, W., Nota, A.,
Jedrzejewski, R.I., \&\ Paresce, F. 1993, in Space Astronomical
Telescopes and Instruments II, edited by P.Y.\ B\'ely \&\ J.B.\
Breckinridge, (SPIE, Spellingham), p.\ 252
\reference Baxter, D.A. 1994a, Instrument Science Report FOC-073,
(STScI, Baltimore)
\reference Baxter, D.A. 1994b, in Calibrating Hubble Space
Telescope, edited by J.C.\ Blades, \&\ S.J.\ Osmer, (STScI,
Baltimore), p.\ 109
\reference Bergvall, N. 1985, \aap, 146, 269
\reference Binggeli, B., Tammann, G.A., \&\ Sandage, A.  1987, \aj,
94, 251
\reference Bruzual, A.G., \&\ Charlot, S. 1993, \apj, 405, 538
\reference Buat, V., Donas, J., \&\ Deharveng, J.M. 1987, \aap, 185, 33
\reference Burstein, D., \&\ Heiles, C. 1978, \apj, 225, 40
\reference Burstein, D., \&\ Heiles, C. 1982, \aj, 87, 1165
\reference Burstein, D., \&\ Heiles, C. 1984, \apjs, 54, 33
\reference Caldwell, N., \&\ Phillips, M.M. 1989, \apj, 338, 789
\reference Calzetti, D., Kinney, A.L., \&\ Storchi-Bergmann, T. 1994,
\apj, 429, 582 (C94)
\reference Calzetti, D., Bohlin, R.C., Kinney, A.L., Storchi-Bergmann,
T., \&\ Heckman, T.M. 1995, \apj, 443, 136
\reference Chen, P.C., Cornett, R.H., Roberts, M.S., Bohlin, R.C.,
Neff, S.G., O'Connell, R.W., Parise, R.A., Smith, A.M., \&\ Stecher,
T.P. 1992, \apjl, 395, L41
\reference Conti, P.S., \&\ Vacca, W.D. 1994, \apjl, 423, L97
\reference Courvoisier, T.J.-L., Reichen, M., Golay, M., \&\ Huguenin,
D. 1990, \aap, 238, 63
\reference Davidson, K., Kinman, T.D., \&\ Friedman, S.D. 1989, \aj,
97, 1591
\reference Deharveng, J.-M., Sasseen, T.P., Buat, V., Lampton, M., and
Wu, X. 1994, \aap, 289, 715
\reference de Vaucouleurs, G., de Vaucouleurs, A.,  Corwin, H.G.,
Buta, R.J., Paturel, G., \&\ Fouqu\'e, P. 1991, Third Reference
Catlog of Bright Galaxies, (Springer-Verlag, New York), (RC3)
\reference D\'esert, F.-X., Boulanger, \&\ Puget, J.L.  1990, \aap,
237, 215
\reference Donas, J., Deharveng, J.M., Laget, M., Milliard, B.,
Huguenin, D. 1987, \aap, 180, 12
\reference Doyon, R., Puxley, P.J., \&\ Joseph, R.D.  1992, \apj, 397, 117
\reference Draine, B.T., \&\ Salpeter, E.E. 1979, \apj, 231, 77
\reference Dufour, R.J., Garnett, D.R., \&\ Shields, G.A. 1988, \apj,
332, 752
\reference Dufour, R.J., \&\ Hester, J.J. 1990, \apj, 350, 149
\reference Elson, R.A.W., \&\ Fall, S.M.  1985, \pasp, 97, 692
\reference Fall, S.M., \&\ Rees, M.J. 1977, \mnras, 181, 37P
\reference Fall, S.M., \&\ Rees, M.J. 1988, in
The Harlow-Shapley Symposium on Globular Cluster Systems in
Galaxies (IAU symposium 126), edited by J.E.\ Grindlay \&\ A.G.\
Davis Philip, (Kluwer, Dordrecht), p.\ 323
\reference Fanelli, M.N., O'Connell, R.W., \&\ Thuan, T.X. 1988, \apj,
334, 665
\reference Feinstein, C., Vega, I., Mendez, M., \&\ Forte, J.C. 1990,
\aap, 239, 90
\reference Fitzpatrick, E.L. 1985, \apj, 294, 219
\reference Forbes, D.A., Norris, R.P., Williger, G.M., \&\ Smith, R.C.
1994a, \aj, 107, 984 (F94a)
\reference Forbes, D.A., Kotilainen, J.K., \&\ Moorwood, A.F.M.
1994, \apj, 433, L13 (F94b)
\reference Forti, G., 1993, IAU Circular, 5719
\reference Friedman, S.D., Cohen, R.D., Jones, B., Smith, H.E., and
Stein, W.A. 1987, \aj, 94, 1480
\reference Garcia, A.M. 1993, \aaps, 100, 47
\reference Gehrz, R.D., Sramek, R.A., \&\ Weedman, D.W. 1983, \apj,
267, 551
\reference Graham, J.A. 1981, \pasp, 93, 552
\reference Greenfield, P., Nota, A., Jedrzejewski, R., Hack, W.,
Hasan, H., Hodge, P., Baxter, D., Baggett, W., \&\ Paresce, F. 1993,
in Space Astronomical Telescopes and Instruments II, edited by P.Y.\
B\'ely \&\ J.B.\ Breckinridge, (SPIE, Spellingham), p.\ 264
\reference Greenfield, P. 1994, Instrument Science Report FOC-075,
(STScI, Baltimore)
\reference Grothues, H.G., \&\ Schmidt-Kaler, T. 1991, \aap, 242, 357
\reference Hack, W. \&\ Nota, A. 1994, Instrument Science Report FOC-080,
(STScI, Baltimore)
\reference Harris, W.E. 1991, \araa, 29, 543
\reference Heckman, T.M., Armus, L., \&\ Miley, G.K. 1990, \apjs, 74, 833
\reference Heckman, T.M. 1994, in {\em Mass-Transfer Induced Activity
in Galaxies\/}, edited by I.\ Shlossman, (University of Cambridge
Press, Cambridge), p.\ 234
\reference Helou, G., Khan, I.R., Malek, L., \&\ Boehmer, L. 1988,
\apjs, 68, 151
\reference Hill, J.K., Bohlin, R.C., Cheng, K.-P., Hintzen, P.M.N.,
Landsman, W.B., Neff, S.G., O'Connell, R.W., Roberts, M.S. Smith, A.M.
Smith, E.P. \&\ Stecher, T.P. 1992, \apjl, 395, L37
\reference Holtzman, J., et al.\ (the WFPC team) 1992, \aj, 103, 691
\reference Howarth, I.D. 1983, \mnras, 203, 301
\reference Hunter, D.A., O'Connell, R.W., \&\ Gallagher, J.S. 1994a,
\aj, 108, 84
\reference Hunter, D.A., van Woerden, H., \&\ Gallagher, J. 1994b,
\apjs, 91, 79
\reference Iye, M., Ulrich, M.-H., \&\ Peimbert, M. 1987, \aap, 186, 84
\reference Joy, M., Lester, D.F., Harvey, P.M., Telesco, C.M., Decher,
R., Rickard, L.J., \&\ Bushouse, H. 1989, \apj, 339, 100
\reference Keel, W.C., Kennicutt, R.C., Hummel, E., \&\ van der Hulst,
J.M. 1985, \aj, 90, 708
\reference Kennicutt, R.C., \&\ Kent, S.M. 1983, \aj, 88, 1094
\reference Kennicutt, R.C. 1989, \apj, 344, 685
\reference Kennicutt, R.C., Edgar, B.K., \&\ Hodge, P.W. 1989, \apj,
337, 761
\reference King, I.R. 1966, \aj, 71, 64
\reference Kinney, A.L., Bohlin, R.C., Calzetti, D., Panagia, N., and
Wyse, R.F.G. 1993, \apjs, 86, 5 (K93)
\reference Kinney, A.L., Calzetti, D., Bica, E., \&\ Storchi-Bergmann,
T. 1994, \apj, 429, 172 (K94)
\reference Kunth, D., Lequeux, J., Sargent, W.L.W., \&\ Viallefond,
F. 1994, \aap, 709, 282
\reference Larson, R.\ 1988, in The Harlow-Shapley Symposium on Globular
Cluster Systems in Galaxies (IAU symposium 126), edited by J.E.\ Grindlay
and A.G.\ Davis Philip, (Kluwer, Dordrecht), p.\ 311
\reference Larson, R. 1993,  in The Globular
Cluster-Galaxy Connection (Astron.\ Soc.\ of the Pacific Conf.\ Series
Vol.\ 48), edited by G.H.\ Smith \&\ J.P.\ Brodie, (ASP, San
Fransisco), p.\ 675
\reference Lasker, B.M., Sturch, C.R., McLean, B.J., Russell J.L.,
Jenkner, H., \&\ Shara, M.M. 1990, \aj, 99, 2059
\reference Lehnert, M. 1992, Ph.D. Thesis, The Johns Hopkins University
\reference Leitherer, C., \&\ Heckman, T.M. 1995, \apjs, 96, 9 (LH)
\reference Lucy, L.B. 1974, \aj, 79, 745
\reference Malumuth, E.M., \&\ Heap, S.R. 1994, \aj, 107, 1054
\reference Marlowe, A.T., Heckman, T.M., Wyse, R.F.G., \&\ Schommer,
R. 1995, \apj, 438, 563
\reference Mathis, J.S. 1994, \apj, 422, 176
\reference Mazzarella, J.M., \&\ Boroson, T.A. 1993, \apjs, 85, 27
\reference McLeod, K.K., Rieke, G.H., Rieke, M.J., \&\ Kelly, D.M.
1993, \apj, 412, 111
\reference Melnick, J., Moles, M., \&\ Terlevich, R. 1985, \aap,
149, L24
\reference Meurer, G.R. 1989, Ph.D. Thesis, Australian National University
\reference Meurer, G.R., Freeman, K.C., Dopita, M.A., \&\ Cacciari, C.
1992, \aj, 103, 60 (MFDC)
\reference Meurer, G.R. 1994, in ESO/OHP Workshop on Dwarf
Galaxies (ESO Conference \&\ Workshop Proceedings no.\ 49), edited
by G.\ Meylan, \&\ P.\ Prugniel, (ESO, Garching), p.\ 351
\reference Meurer, G.R., Mackie, G., \&\ Carignan, C. 1994, \aj, 107, 2021
\reference Meurer, G.R. 1995a Instrument Science Report FOC-083,
(STScI, Baltimore)
\reference Meurer, G.R. 1995b, Nature, 375, 742
\reference Moshir, M., Kopan, G., Conrow, T., McCallon, H., Hacking, P.,
Gregorich, D., Rohrbach, G., Melnyk, M., Rice, W., Fullmer, L.,
et al.\ 1990 Infrared Astronomical Satellite Catalogs, The Faint Source
Catalog, Version 2.0
\reference Nota, A., Jedrzejewski, R., \&\ Hack, W.\ (editors) 1993,
Hubble Space Telescope Faint Object Camera Instrument Handbook
(version 4.0), (STScI, Baltimore)
\reference O'Connell, R.W., Gallagher, J.S., \&\ Hunter, D.A., 1994,
\apj, 433, 65, 1994
\reference O'Connell, R.W., Gallagher, J.S., \&\ Hunter, D.A., and
Colley, W.N. 1995, \apjl, 446, L1
\reference Osterbrock, D.E., \&\ Shaw, R.A. 1988, \apj, 327, 89
\reference Panagia, N. 1977, in Infrared Astronomy, edited by G.\
Setti \&\ G.G.\ Fazio, (Reidel, Dordrecht), p.\ 115
\reference Pastoriza, M.G., Dottori, H.A., Terlevich, E., Terlevich,
R., \&\ D\'iaz, A.I.  1993, \mnras, 260, 177
\reference Peterson, C.J., \&\ King, I.R. 1975, \aj, 80, 427
\reference Pettini, M., \&\ Lipman, K. 1995, \aap, 297, L63
\reference Reichen, M., Kaufman, M., Blecha, A., Golay, M. and
Huguenin, D. 1994, \aaps, 106, 523
\reference Richardson, W.H. 1972, Opt.Soc.Am., 62, 55
\reference Rieke, G.H., Lebofsky, M.J, Thompson, R.I., Low, F.J., and
Tokunaga, A.T. 1980 \apj, 238, 24
\reference Rieke, G.H., Loken, K., Rieke, M.J., \&\ Tamblyn, P. 1993,
\apj, 412, 99
\reference Robert, C., Heckman, T.M., Leitherer, C., Garnett, D.R.,
Kinney, A., van de Rydt, F., \&\ Meurer, G.R. 1995, {\em in preparation}
\reference Russell J.L., Lasker, B.M., McLean, B.J., Sturch, C.R., and
Jenkner, H. 1990, \aj, 99, 2059
\reference Saha, A., Sandage, A., Tammann, G.A., Labhardt, L.,
Schwengeler, H., Panagia, N., Macchetto, F.D. 1995, \apj, 438, 8
\reference Salpeter, E.E. 1955, \apj, 121, 161
\reference Sandage, A., Saha, A., Tamman, G.A., Labhardt, L.,
Schwengeler, H., Panagia, N., \&\ Macchetto, F.P.  1994, \apjl, 423, L13
\reference Schechter, P.L. 1980, \aj, 85, 801
\reference Searle, L., \&\ Zinn, R. 1978, \apj, 225, 357
\reference Seaton, M.J. 1979, \mnras, 187, 73p
\reference Skillman, E.D. \&\ Kennicutt, R.C. 1993, \apj, 411, 655
\reference Soifer, B.T., Sanders, D.B., Madore, B.F., Neugebauer, G.,
Danielson, G.E., Elias, J.H., Lonsdale, C.J., \&\ Rice, W.L.  1987,
\apj, 320, 238
\reference Sparks, W.B.  1991, Instrument Science Report FOC-053,
(STScI, Baltimore)
\reference Stanford, S.A., \&\ Wood, D.O.S. 1989, \apj, 346,
712
\reference Stetson, P.B. 1987, \pasp, 99, 191
\reference Stetson, P.B. 1993,  in The Globular Cluster-Galaxy
Connection (Astron.\ Soc.\ of the Pacific Conf.\ Series Vol.\ 48),
edited by G.H.\ Smith \&\ J.P.\ Brodie, (ASP, San Fransisco), p.\ 14
\reference Telesco, C.M., \&\ Gatley, I. 1984, \apj, 284, 557
\reference Thuan, T.X. \&\ Martin G.E. 1981, \apj, 247, 823
\reference Toomre, A. 1977, in The Evolution Of Galaxies And
Stellar Populations, edited by B.\ Tinsley \&\ R. Larson (Yale
University Observatory, New Haven), p.\ 401
\reference Treffers, R. R., Leibundgut, B., \&\ Filippenko, A. V.
1993, IAU Circular, 5718
\reference van Buren, D., Jarrett, T., Terebey, S., \&\ Beichman, C.
1994, IAU Circular, 5960
\reference van den Bergh, S., Morbey, C., \&\ Pazder, J.  1991, \apj,
375, 594
\reference van den Bergh, S. 1995, Nature, 374, 215
\reference van der Kruit, P.C., \&\ de Bruyn, A.G. 1976, \aap, 48, 373
\reference van der Kruit, P.C. 1976, \aap, 49, 161
\reference Walsh, J.R., \&\ Roy, J.-R.  1987, \apjl, 319, L57
\reference Walsh, J.R., \&\ Roy, J.-R.  1989, \mnras, 239, 297
\reference Weedman, D.W., 1988, Astro.\ Lett.\ \&\ Comm., 27, 117
\reference Weedman, D.W., 1991, in Massive Stars in
Starbursts, edited by C.\ Leitherer, N.R.\ Walborn, T.M.\ Heckman, \&\
C.A.\ Norman, (Cambridge University Press, Cambridge), p.\ 317
\reference Whitmore, B.C., Schweizer, F., Leitherer, C., Borne, K.,
\&\ Robert, C. 1993, \aj, 106, 1354
\reference Whitmore, B.C., \&\ Schweizer, F. 1995, \aj, 109, 960 (WS)
\reference Witt, A.N., Thronson, H.A., Capuano J.M.  1992, \apj, 393, 611
\reference Wynn-Williams, C.G., Eales, S.A., Becklin, E.E., Hodapp,
K.W., Joseph, R.D., McLean, I.S., Simons, D.A., \&\ Wright, G.S. 1991,
\apj, 377, 426
\end{references}
\end{document}